# Stabilization of $LaH_{10}$ with Thorium: Superconductivity in $(La,Th)H_{10}$ up to 232 K near 1 Mbar

Dmitrii V. Semenok[1,†,*], Mikhail A. Volkov[2,†], Andrey V. Sadakov[3,†], Alexander T. Daniyarkhodzhaev[3], Alexey S. Usoltsev[3], Vladimir M. Pudalov[3], Mikhail I. Volgin[2], Iurii M. Nevolin[2], Toni Helm[4], Di Zhou[1], Viktor V. Struzhkin[5,*], and Ivan A. Troyan[6,*]

[1] *Center for High Pressure Science and Technology Advanced Research (HPSTAR), Beijing*

[2] *Institute of Physical Chemistry and Electrochemistry of the Russian Academy of Sciences, 119071, Moscow, Leninsky Prospekt, 31/4, Russia*

[3] *V. L. Ginzburg Center for High-Temperature Superconductivity and Quantum Materials, Moscow, 119991 Russia*

[4] *Dresden High Magnetic Field Laboratory (HLD-EMFL), Helmholtz-Zentrum Dresden-Rossendorf (HZDR), 01328 Dresden, Germany*

[5] *Center for High Pressure Science and Technology Advanced Research (HPSTAR), Shanghai*

[6] *Shanghai Key Laboratory of Material Frontiers Research in Extreme Environments (MFree), Shanghai Advanced Research in Physical Sciences (SHARPS), Pudong, Shanghai 201203, China*

*Corresponding authors, emails: dmitrii.semenok@hpstar.ac.cn (Dmitrii V. Semenok), viktor.struzhkin@hpstar.ac.cn (Viktor V. Struzhkin), and itrojan@sharps.ac.cn (Ivan A. Troyan)

†These authors contributed equally to this work

## Abstract

Ternary lanthanum-based hydrides are among the best high-temperature superconductors. Here we show that substituting La with thorium stabilizes $La(Th)H_{10}$ clathrate structure at ~1 Mbar while only minimally suppressing superconductivity. The resulting cubic $(La,Th)H_{10}$ phase, synthesized from relatively air-stable $La_2Th$, $La_3Th$ and $La_5Th$ alloys, reaches onset $T_c$ = 232 K at 134 GPa, as confirmed by contact resistance measurements. Superconductivity was verified by two independent methods: electrical resistance and contactless radio-frequency measurements under applied magnetic fields at various pressures. Using a record-high pulsed current of 380 mA, we investigated critical phenomena in $P6_3/mmc$-$(La,Th)H_9$ at 173 GPa, which allowed for the experimental determination of a single superconducting gap within s-wave model: $\Delta(0)$ = 37 ± 4.5 meV and $2\Delta(0)/kT_c$ = 4.9 ± 0.6. The La-Th-H system is chemically rich: in addition to the decahydride and nonahydride mentioned above, we also identified nanocrystalline $I4/mmm$-$(La,Th)H_4$ (5-6 nm grain size), $(La,Th)H_6$, as well as $hP13$, $cF80$ and A15 phases with compositions of $(La,Th)H_{5.75\text{-}6.25}$ and $T_c \approx$ 83 K at 70-80 GPa, and a possible $(La,Th)H_{16}$ polyhydride.

## 1. Introduction

The exploration of high-temperature superconductivity in hydrides has emerged as a transformative frontier in condensed matter physics, driven by the discovery of superconducting binary hydrides such as $H_3S$ [1], $LaH_{10}$ [2,3], $YH_6$ [4], $ThH_{9\text{-}10}$ [5] under extreme pressures [6]. These materials exhibit remarkable superconducting critical temperatures ($T_c$): for instance, thorium polyhydrides $ThH_9$ and $ThH_{10}$ demonstrate $T_c$ of 143 K and 159-161 K, respectively, at pressures above 150 GPa [5]. Similarly, lanthanum decahydride $LaH_{10}$ has been shown to achieve a $T_c$ of at least 250 K at similar pressure conditions, making it one of the highest-$T_c$ superconductors known to date [2,3]. However, stabilizing these phases often requires ultrahigh pressures, limiting their practical applicability [7]. Recent advancements have demonstrated that the incorporation of cerium into the La lattice of the $LaH_{10}$, forming ternary La-Ce polyhydrides with a solid solution sublattice, can stabilize the clathrate cubic superconducting phase $XH_{10}$ at significantly lower pressures of 95-130 GPa while maintaining a high maximum $T_c$ of 180-190 K [8,9]. This stabilization is attributed to the $CeH_{9\text{-}10}$ clusters, existing in the (La,Ce)$H_{9\text{-}10}$ structure, which themselves demonstrate increased thermodynamic and dynamic stability[10]. Approximately the same result is achieved in the quaternary hydride involving cerium, (La,Y,Ce)$H_{9\text{-}10}$, for which $T_c$ also reaches 190 K at a remarkably low pressure of 112 GPa [11]. The point is that Ce atom has an extra *4f* electron and, by virtue of its higher valence, binds the hydrogen framework into a unified whole more effectively than the La atom. However, cerium also has a negative effect: due to the presence of a magnetic moment, cerium ions scatter singlet Cooper pairs with a spin flip, thereby suppressing conventional superconductivity, which leads to a more than two-fold decrease in the critical temperature of $CeH_{10}$ compared to $LaH_{10}$ (115 K [10] versus 250 K [2,3]), analogously to what is observed for Nd-doped superhydrides [12,13].

Despite promising results, obtained for (La,Ce)$H_{9\text{-}10}$, the quest for further enhancement of $T_c$ and decreasing of the stabilization pressure remains a central challenge in the field. One encouraging strategy involves replacing cerium, an *f*-electron element, with thorium, a *d*-electron one. This substitution is expected to modify the electronic structure and electron–phonon coupling, potentially enhancing superconductivity. Preliminary theoretical studies of La-Th-H system suggest that thorium incorporation into lanthanum hydrides could elevate the critical temperature by an additional 20-30 K, offering a pathway to achieve $T_c$ values exceeding 210 K in the same pressure range of around 1 Mbar [14]. The recently investigated Y-Th-H system does show a definite increase in $T_c$ to 184 K compared to 161 K in pure thorium hydrides [15]. This approach is in full agreement with the previously put forward idea of the localization of superconducting hydrides in the *d*-belt of the Mendeleev Table and the negative influence of *f*-electrons on superconductivity in them [16].

In this work, we investigate the crystal structure, electronic properties, and superconductivity of lanthanum-thorium ternary hydrides, aiming to preserve the most pronounced superconductivity possible at the lowest pressure. Using various La-Th alloys and ammonia borane ($NH_3BH_3$), we synthesized a set of La-Th superhydrides with a mixed solid-solution sublattice, stable below 100 GPa and exhibiting superconductivity with $T_c$ up to 232 K.

## 2. Results

DACs S0, and S5 were prepared for X-ray diffraction (XRD) studies at low pressures (< 130 GPa) and did not contain electrodes. DACs S1, S2 were prepared to study the superconducting properties of La-Th hydrides at pressures of 100-140 GPa. DACs S3 and S4 were prepared to study the high-pressure region (173-180 GPa). The pulsed field DAC was made to study the behavior of superconductivity in La-Th hydrides in strong pulsed magnetic fields, but due to a crack in one of the electrodes, this task was not completed. The DAC SM (matrix) was prepared for radio-frequency studies and to check the effect of excess of ammonia borane (AB) on the synthesis results. The sample in this DAC had the largest area. The culets of the diamond anvils in all DACs had a diameter of 50 μm, except for DACs SM and S5, where the culet had a diameter of 100 and 75 μm, respectively.

In all experiments, sublimed ammonia borane ($NH_3BH_3$) served as the internal hydrogen source. High-pressure laser heating was performed using 50–300 ms short pulses from a 1-μm infrared fiber laser (IPG Photonics). We examined three La-Th alloys prepared by melting the lanthanum together with the thorium in an electric arc furnace: $La_2Th$, $La_3Th$, and $La_5Th$. Detailed information on the samples studied is provided in Supporting Table S1.

### *2.1 Synthesis and X-ray diffraction studies of $XH_9$ and $XH_{10}$ phases*

From the structural point of view, the behavior of the La-Th-H system under pressure is similar to La-H [2,3,17] and Th-H [5,18] systems: the main synthesis products are cubic *fcc* decahydride (La,Th)$H_{10}$ and hexagonal nonahydride $P6_3/mmc$-(La,Th)$H_9$ with mixed metal sublattices. DAC S0 was used to study stability of the La-Th polyhydrides during decompression from 128 GPa. Electrical DAC S3 with Au/Ta electrodes was studied via X-ray diffraction (XRD) to identify phases responsible for the superconductivity at pressure of 173 GPa. At both pressures, 128 and 176 GPa, the $P6_3/mmc$-$XH_9$ (on edges of laser heated spots, X = (La,Th)) and cubic $Fm\bar{3}m$-$XH_{10}$ (in the center of the laser spot) are formed as the main products (Figure 1a). The sample also contains minor phases, possibly pseudohexagonal $Cmc2_1$-(La,Th)$H_6$, as well as some other phases giving uninterpreted XRD peaks (asterisks, Figures 1a,c). The interpretation of these peaks is difficult due to a lack of sufficient information and overlapping with strong reflections from (La,Th)$H_9$ and (La,Th)$H_{10}$. The unit cell volumes of the latter phases are very close to those of $LaH_{10}$ [2,3], $ThH_{10}$ and $ThH_9$ [5] (Figure 1e).

The cubic $XH_{10}$ phase begins to decompose around 93-95 GPa, where refining the structure can no longer be performed with sufficient accuracy using the ideal *fcc* structure, and also the volume of the unit cell drops due to the loss of hydrogen (Figure 1e). The hexagonal $XH_9$ is more stable and decomposes at pressures below 90 GPa (Figure 1e). Due to the cracking of the diamond anvil and the consequent rapid decompression down to 60 GPa, we were unable to establish an exact boundary of the stability region of (La,Th)$H_9$. Additional X-ray diffraction data for the synthesized hydrides are provided in the Supporting Information.

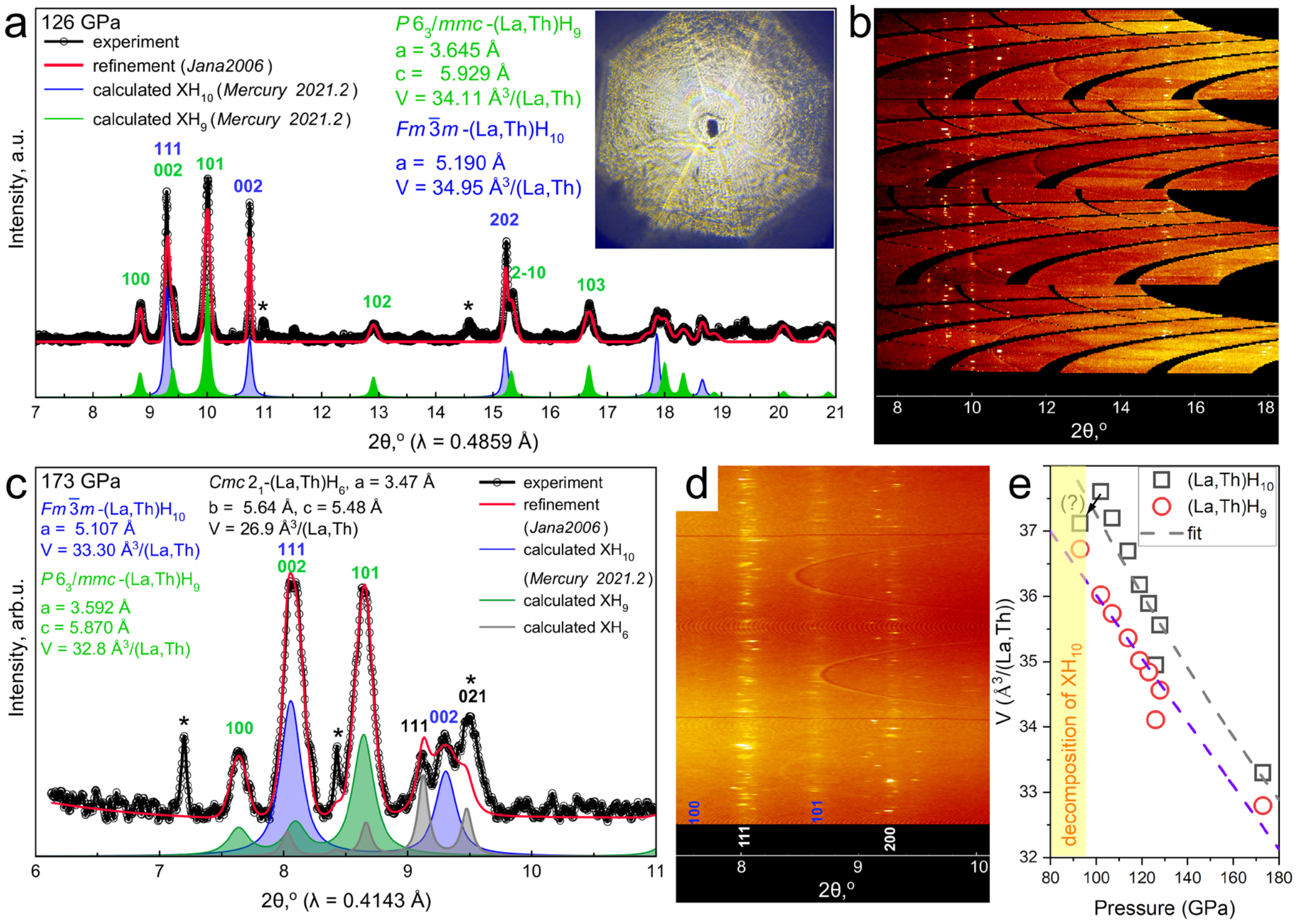

**Figure 1.** Experimental X-ray diffraction pattern (black points) of (La,Th, 2:1) hydrides at 126 and 173 GPa. (a) The red line shows the Le Bail refinement fit, incorporating both hexagonal (La,Th)$H_9$ and cubic (La,Th)$H_{10}$ phases. Theoretically calculated (Mercury 2021.2 software [19]) diffraction patterns for hexagonal (La,Th)$H_9$ and cubic (La,Th)$H_{10}$ are shown in green and blue, respectively, along with their refined lattice parameters. Inset: Photograph of the high-pressure diamond anvil cell with the (La,Th)$H_x$ sample. (b) Original 2D diffraction image ("cake") from which the 1D pattern was integrated. (c) X-ray diffraction pattern and Le Bail refinements of $XH_{10}$, $XH_9$, and $XH_6$ unit cells of the La-Th hydrides at 173 GPa, the electrical DAC S3. Data were collected at $\lambda = 0.4143$ Å and refined using Jana2006 and Mercury 2021.2. Asterix marks (*) correspond to unidentified La-Th hydrides. (d) Original 2D diffraction image ("cake") from which the 1D pattern was integrated. (e) Experimental unit cell volumes per (La,Th) as a function of pressure for $XH_{10}$ and $XH_9$ phases, measured during decompression of DAC S0. Yellow area and (?) sign indicate region of $XH_{10}$ decomposition. Data points at 173 GPa belong to DAC S3.

### *2.2 Small-quantity La–Th polyhydrides: bct-$XH_4$, cF80, hP13 and A15 phases*

In addition to the main phases $XH_{10}$ and $XH_9$, a number of by-products can also be observed after laser heating of La-Th alloys with AB. For example, in DAC S1 we obtained *P6/mmm*-$X_3H_{11-x}$ (= $XH_{3.7}$, Figures 2a,b) at 126 GPa, known from studies of phase transformations of Y, Nd and Gd hydrides at high pressure[20,21], and prone to non-stoichiometric compositions. Although other structures with *P*6/*mmm* space group can explain the presence of additional diffraction peaks (Supporting Figure S15), it is $X_3H_{11}$ that best describes the diffraction pattern. This lower hydride with atomic H-sublattice is of no interest from the point of view of superconductivity.

Another well-known hydride, the tetragonal $XH_4$, was synthesized in the DAC S4 at 174–180 GPa (Figure 2c) due to an unforeseen incident during a wide-focus $CO_2$ laser heating. During laser heating, the edge of the focus caught on the $CaF_2$/epoxy insulating gasket, causing it to lose its viscosity and be forced out of the vicinity of the culet within milliseconds or microseconds, thereby also damaging the Au/Ta electrodes. This shock compression of the sample led to a pressure increase from an initial 174 GPa to 180 GPa, and to the grain refinement of *I*4/*mmm*-(La,Th)$H_4$ to a size of 5–6 nm (Scherrer[22] and Williamson[23], FWHM = 0.4–0.5°), consistent with other known examples of hydrides, for instance, A15 $La_4H_{23}$ [24,25]. Internal stresses are found to be practically zero, and instrumental broadening is negligible (FWHM($CeO_2$) ≈ 0.029°). This example is interesting in that the resulting sample is completely homogeneous and single-phase (see Supporting Figure S16), while the diffraction rings are very diffuse. Thus, this experiment demonstrates the feasibility of synthesizing homogeneous, single-phase nanocrystalline superhydride samples.

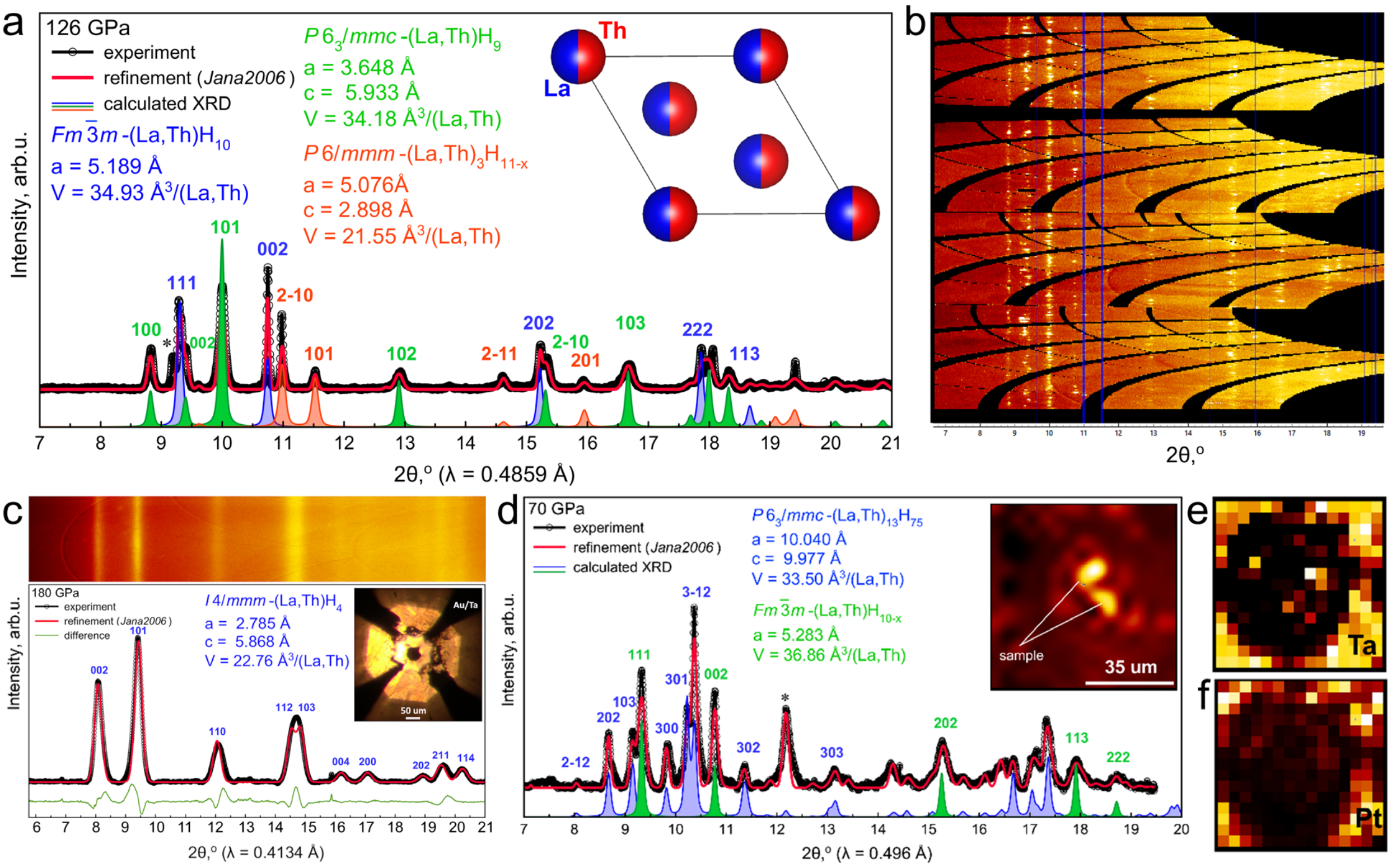

**Figure 2.** Experimental XRD and the Le Bail refinements of unit cells of secondary La-Th hydrides in DACs S1, S4, and S5. (a) The integrated 1D experimental profile of DAC S1's sample at 126 GPa shown as a black line with circles, with the Le Bail refinement (Jana2006) of unit cell parameters of $XH_{10}$, $XH_9$, and $X_3H_{11}$ phases indicated by the red line. Calculated XRD patterns for the *P*6$_3$/*mmc*-(La,Th)$H_9$, cubic (La,Th)$H_{10}$, and *P*6/*mmm*-(La,Th)$_3H_{11-x}$ are shown in green, blue, and orange, respectively. Miller indices (hkl) for each phase are color-coded accordingly, and impurity reflections are marked with asterisks (*). Inset: 2D projection of the $X_3H_{11-x}$ unit cell showing the arrangement of heavy La-Th atoms. (b) Raw 2D diffraction image ("cake") corresponding to the experiment. $X_3H_{11-x}$ solution (blue vertical lines) explains additional small diffraction peak at 9.55 deg. and the peak at 14.6 deg. (c) Experimental XRD pattern (black circle) and Le Bail refinement (red curve) of the unit cell parameters of fine-grained *I*4/*mmm*-(La,Th)$H_4$ in DAC S4 at 180 GPa. Green curve is residual intensity. Inset: optical photo of the sample. Top panel: raw diffraction image ("cake") with diffusive diffraction lines. (d) The same for DAC S5 at 70 GPa. Calculated XRD patterns for the *P*6$_3$/*mmc*-(La,Th)$_{13}H_{75}$, and, possible distorted, cubic (La,Th)$H_{10-x}$, are shown in blue, and green, respectively. Inset: thermal colormap of the sample distribution (10 ± 0.5 deg.) obtained by XRD. (e, f) Spatial distribution of Pt/Ta Lenz lens in DAC S5 near the culet.

The synthesis of superhydrides at pressures below 70 GPa is of great interest. At such pressures, diamond anvils do not crack during decompression and can be reused repeatedly, which significantly reduces the cost of experiments with superhydrides. Using a $La_5Th$/AB sample in the DAC S5 at 70 GPa (Supporting Figure S17), and following the approach of a recent study on A15 $(La,Ce)_4H_{23}$ [26], we synthesized a mixture of La-Th polyhydrides, including the hexagonal *hP*13 $P6_3/mmc$-$(La,Th)_{13}H_{75}$ (=$XH_{5.77}$, Figure 2d), the cubic *cF*80-$(La,Th)_4H_{25}$ (=$XH_{6.25}$, Supporting Figure S18), the cubic A15 $Pm\bar{3}n$-$(La,Th)_4H_{23}$ ($XH_{5.75}$, Supporting Figure S19), and, possibly, the distorted cubic $XH_{10-x}$ which is a product of the decomposition, molecularization and dehydrogenation of *fcc*-$XH_{10}$ below 90 GPa. The *hP*13 and *cF*80 phases were previously discovered by single-crystal diffraction in the Y-H system [21]. Given the high hydrogen content and symmetry in all synthesized $XH_{5-6}$ phases, the sample should exhibit superconducting properties as early as 70 GPa.

Finally, the DAC SM (Supporting Figures S9, S11-S13) exhibits the most interesting and complex chemistry. In addition to the usual *fcc*-$XH_{10}$ phase, it contains the hexagonal $P6_3$/mmc or $P6_3/mc$ $(La,Th)H_6$ and, possibly, a phase with a larger unit cell volume: hexagonal $P6/mmm$-$(La,Th)H_{16}$ (Supporting Figure S11). Previously, we proposed the existence of a hexagonal $P6_3/mc$-$XH_6$ hydride in the Th-H system, along with $Cmc2_1$-$ThH_6$, as a decomposition product of $ThH_{10}$ and $ThH_9$ upon decompression [5,18]. Such structures are close to the convex hull in the Ce-H system [27] but had not previously been detected experimentally. Regarding the second interesting phase, earlier we theoretically investigated a $P6/mmm$-$XH_{16}$-type structure for the La-H system and it was found to be close to the convex hull [28]. A structurally similar $Fmmm$-$CeH_{18}$ and $ThH_{18}$ hydrides were proposed as a promising room-temperature superconductors under ultra-high pressures (350–600 GPa) [29,30]. Interestingly, the *Fmmm* space group provides an even better explanation for the diffraction pattern than $P6/mmm$, but the unit cell volume of this phase better corresponds to the formula $XH_{16}$, rather than $XH_{18}$ (Supporting Figure S13). We will return to the discussion of the presence of these higher hydrides in the DAC SM when discussing superconducting properties.

*2.3 Contact electrical conductivity measurements*

Most La-Th polyhydrides exhibit pronounced high-temperature superconducting properties. In a series of DACs S1-S3, and in the pulsed field DAC, loaded with $La_2Th$ and $La_3Th$ alloys, we investigated the dependence of $T_c$ on the synthesis pressure (Figures 3a, 5a). We found that laser heating of $La_2Th$ at a pressure of about 100 GPa leads to the formation of superconducting $(La,Th)H_{10}$ with a cubic structure and $T_c$(onset) above 190 K. Increasing the synthesis pressure to 134 GPa allows reaching a maximum of $T_c$(onset) ≈ 232 K, and then the critical temperature slightly decreases with increasing pressure, demonstrating an asymmetric bell-shaped dependence of $T_c(P)$ (Figures 5a, c), which has been observed many times for other superhydrides [1,2,10].

The synthesized $(La,Th)H_{9-10}$ samples in all cases demonstrate near-zero residual resistance (for example, $R_{SC}$ ≈ 15 μΩ, see inset in Figure 3a) in the main channels of the van der Pauw scheme. Superconducting transitions are linearly shifted in a magnetic field to the lower temperature region ($dT_c/dB$ = – 1.65 K/T, Figures 3b, 5b), demonstrating a visible broadening (Figure 3b). Estimation of coherence length $\xi(0)^2 = \Phi_0/2\pi B_{c2}(0)$ gives 1.65 nm for $(La,Th)H_{10}$, and higher value of 2.73 nm for $(La,Th)H_9$. Expected $B_{c2}(0)$ = 100-120 Tesla.

Using the Bloch-Grüneisen formula, we can estimate the Debye temperature $T_D$ = 1300 K, $\omega_{log} \approx 0.827 \times T_D$ = 1075 K and the electron-phonon interaction constant $\lambda \approx 2.45$ at 173 GPa in DAC S3 using the inversion of the Allen-Dynes formula [31] with $\mu^*$=0.1. The obtained estimates are in agreement with first-principles calculations ($\lambda_{DFT}$ = 2.17–2.55, Supporting Figure S29). At the same time, fitting the normal part of the electrical resistance of the DAC S3 sample at 173 GPa indicates the presence of anomalies in the vicinity of 238-250 K (Supporting Figures S22, S25), near the maximum $T_c$ predicted for La-Th-H system [14], which are better expressed in the plot of the first derivative of resistance with respect to temperature $dR/dT$ (Supporting Figure S22b). This anomaly may correspond to the appearance of superconductivity in impurity phases, such as $XH_{16}$, present in the sample in a small amount.

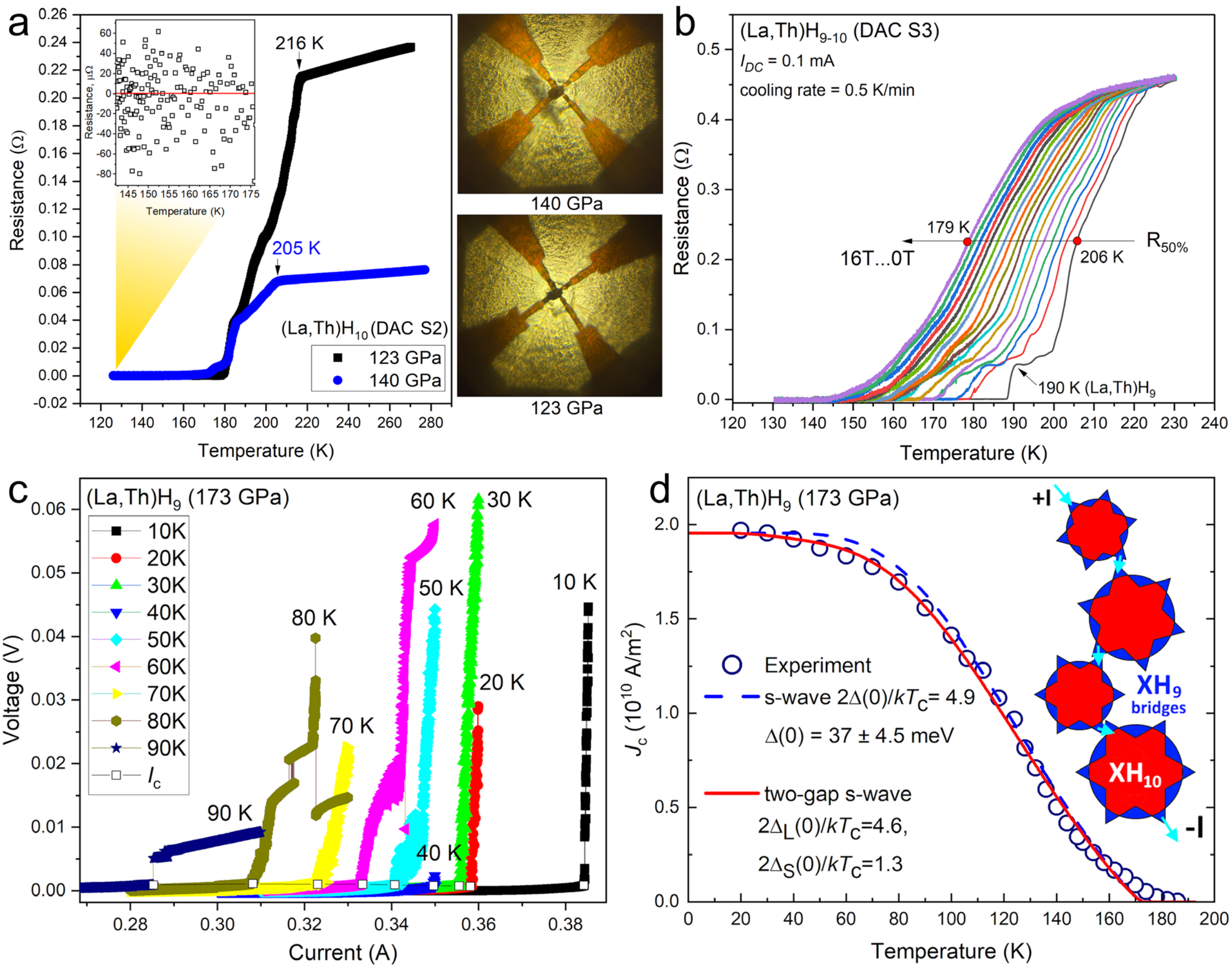

**Figure 3.** Superconducting properties of La-Th polyhydrides at different pressures. (a) Temperature dependence of resistance for (La,Th)$H_{9-10}$ at two different pressures (DAC S2, 123 GPa and 140 GPa), showing their onset critical temperatures $T_c$(onset) at 216 K and 205 K. The left inset displays the residual resistance. The right insets show optical images of samples at 140 GPa (top) and 123 GPa (bottom). (b) Resistance as a function of temperature in magnetic fields from 0 to 16 T for the DAC S3 sample. A small step at 190 K is observed, attributed to the presence of the hexagonal $XH_9$ phase. $T_c$ was calculated using the $R_{50\%}$ criterion. (c) Current-voltage characteristics of the DAC S3 sample at 173 GPa, more precisely, of the (La,Th)$H_9$ bridge, in the range from 90 K to 10 K measured using a dedicated pulsed current source up to 380 mA. (d) Dependence of the critical current density (cross section: 2×7 μm$^2$) on temperature for DAC S3. Interpolation is performed using the one- and two-gap s-wave superconductivity models [32]. Inset: proposed trajectory of current through thin hexagonal $XH_9$ bridges connecting cubic $XH_{10}$ grains.

In order to evaluate the superconducting gap, we have studied the current-voltage characteristics of the DAC S3 sample using a pulsed current source of up to 380 mA (Figure 3c), which allows suppressing the superconducting $T_c$ down to 10 K in the hexagonal (La,Th)$H_9$ bridges with $T_c$(offset) = 175 ± 2 K at 173 GPa. Interpolation of the experimental data on the temperature dependence of critical current $I_c(T)$ and critical current density $J_c(T)$, calculated from the cross section of the sample near the Ta/Au electrode (width: 7 μm, thickness: 2 μm), yields the following parameters for s-wave superconductivity: $\Delta(0)$ = 37 ± 4.5 meV, $2\Delta(0)/kT_c$ = 4.9 ± 0.6, the heat capacity jump $\Delta C/C$ = 0.8 ± 0.1, and the London penetration depth $\lambda_L(0) \leq 220$ ± 20 nm (upper estimate, Figure 3d). Corresponding Ginzburg-Landau parameter $\kappa(0) = \lambda_L(0)/\xi(0) \leq 80$ for (La,Th)$H_9$ at 173 GPa. Since the experimental critical current in polycrystalline samples is always underestimated due to intergranular barriers, estimates based on $\lambda_L$ and $\kappa$ should be understood as upper limits. The same coherence length and the superconducting gap (Figure 3d) leads to a reasonable value of the Fermi velocity $V_F = \pi\Delta(0)\xi(0)/\hbar$ = 4.8×10$^5$ m/s, typical for superhydrides [33]. All the main parameters of the superconducting state in La-Th hydrides are collected in Supporting Table S3.

The obtained value of the superconducting gap $\Delta(0)$ is close to that recently obtained gap for $H_3S$ (≈30 meV [34]) using tunneling spectroscopy with Ta/$Ta_2O_5$ contacts. This is not surprising, since the superconducting gap depends on $T_c$ and logarithmically averaged frequency $\omega_{log}$ = 707–1075 K via, for example, interpolation formula, proposed by Carbotte [35]

$$\frac{2\Delta(0)}{k_B T_c} = 3.53\left(1 + 12.5\left(\frac{T_c}{\omega_{log}}\right)^2 ln\left(\frac{\omega_{log}}{2T_c}\right)\right). \quad (1)$$

Both hydrides, $H_3S$ and (La,Th)$H_9$, have similar $T_c$ and $\omega_{log}$, and eq. 1 gives $\Delta(0)$ = 36–41 meV for (La,Th)$H_9$ and 41 meV for $H_3S$ in an acceptable agreement with the experiment. It is important to note that the penetration depth $\lambda_L(0)$ is comparable with the size of superconducting grains of polyhydrides under pressure, which can smaller than 20–30 nm [36]. This excludes the possibility of complete displacement of the magnetic field from the volume of a microscopic hydride sample (the Meissner effect). Even at 0 K the magnetic field will always penetrate significant part of the sample's volume.

The two-gap alpha model fits the measured $J_c(T)$ data somewhat better when using the following fit parameters: $2\Delta(0)_L/kT_c = 4.6 \pm 0.6$, $\Delta C_L/C_L = 0.9 \pm 0.3$, $2\Delta(0)_S/kT_c = 1.3 \pm 0.3$, $\Delta C_S/C_S = 0.2 \pm 0.1$, $\alpha = 0.9 \pm 0.2$, $\lambda_L(0)$ = 220 ± 20 nm. However, deviation from the single-gap model can also be caused by partial degradation of the sample and a decrease in $J_c(T)$ during measurements (see Supporting Figure S24).

What is the main conclusion that can be drawn from the assessment of superconducting parameters based on critical current density measurements? If we use the data obtained in the critical current experiment for $\lambda_L(0)$ and $\xi(0)$, we can directly estimate the lower critical field for (La,Th)$H_9$ via eq. 2 [37]

$$B_{c1}(0) = \frac{\Phi_0}{4\pi\lambda_L^2(0)}\left[0.497 + \ln\left(\frac{\lambda_L(0)}{\xi(0)}\right)\right], \quad (2)$$

which gives the lower estimate of $B_{c1}(0)$ = 166 Gauss > $B_p(0)$ in qualitative agreement with the $B_{c1}$ of 122-type of Fe-containing superconductors and YBCO [38], but in poor agreement with large $B_{c1}(0)$ values obtained by Minkov et al. [39] using the demagnetization factor for bulk disk-shaped samples. The possible reason of this discrepancy is that for non-uniform superhydride powders, the demagnetization factor $N \approx 0.3$–0.4 and $(1–N)^{-1}$ does not differ much from unity [40]. This has already been emphasized many times in other publications [41-43]. We would like to underline again that for inhomogeneous microscopic samples with grain size $r \leqslant \lambda_L(0)$, the first critical field $B_{c1}$ has no such clear physical meaning as for pure bulk metals, since the magnetic field always penetrates into the sample.

In conclusion of this section, it is important to emphasize that reducing the synthesis pressure in our work is not equivalent to decompressing the DAC with samples synthesized at a higher pressure. Some superhydrides may remain thermodynamically metastable and reach higher values of $T_c$ when decompressed from, say, 150 GPa to 100 GPa at ambient or low temperature (e.g., at 77 K or 4.2 K), but they cannot be synthesized at 100 GPa (for instance, Supporting Figure S23, $T_c$(onset) = 190-195 K) because laser synthesis involves heating to a high temperature ~1000-2000 K, and promotes a fast decomposition of all metastable phases.

Thus, investigating the La-Th-H system, we found that in the pressure range of 100-140 GPa, replacing cerium with thorium in the (La,Ce)$H_{9-10}$ leads to the increase of the onset critical temperature by 20-40 K, up to at least $T_c$ = 232 K.

*2.4 Contactless radio-frequency transmission measurements*

Contactless studies of heterogeneous samples under ultrahigh pressure are an important part of investigating the properties of their components, since the classical four-contact electrode configuration is too crude and provides information about the phases present, at best, only in the form of a series of steps in the resistance versus temperature dependence $R(T)$. Several methods have been proposed for investigating the magnetic response of microscopic samples under high pressure: AC susceptibility [44,45], SQUID magnetometry[39,46], NV centers in diamonds[47], and the radio-frequency detection method (RF)[48-52], which our group is developing. The RF method is an improved and simplified version of the AC susceptibility approach, proposed in 2002 [44], and was already used in 1989 to investigate cuprates in strong pulsed magnetic fields [53]. Despite its widespread use for contactless studies of thin films, we made an important modification to this method for investigations in diamond anvil cells by introducing one or two Lenz lenses[54,55], deposited onto the diamond anvils, into the detection system. Lenz lenses help transmit a sufficient portion of the high-frequency signal directly to the microsample in the culet region of the diamond anvils, substantially increasing the filling factor $V_{sample}/V_{field}$. The objective of this section is to study the behavior of the radio-frequency transmittance of the La-Th hydrides in the vicinity of the resistive superconducting transition (150-300 K).

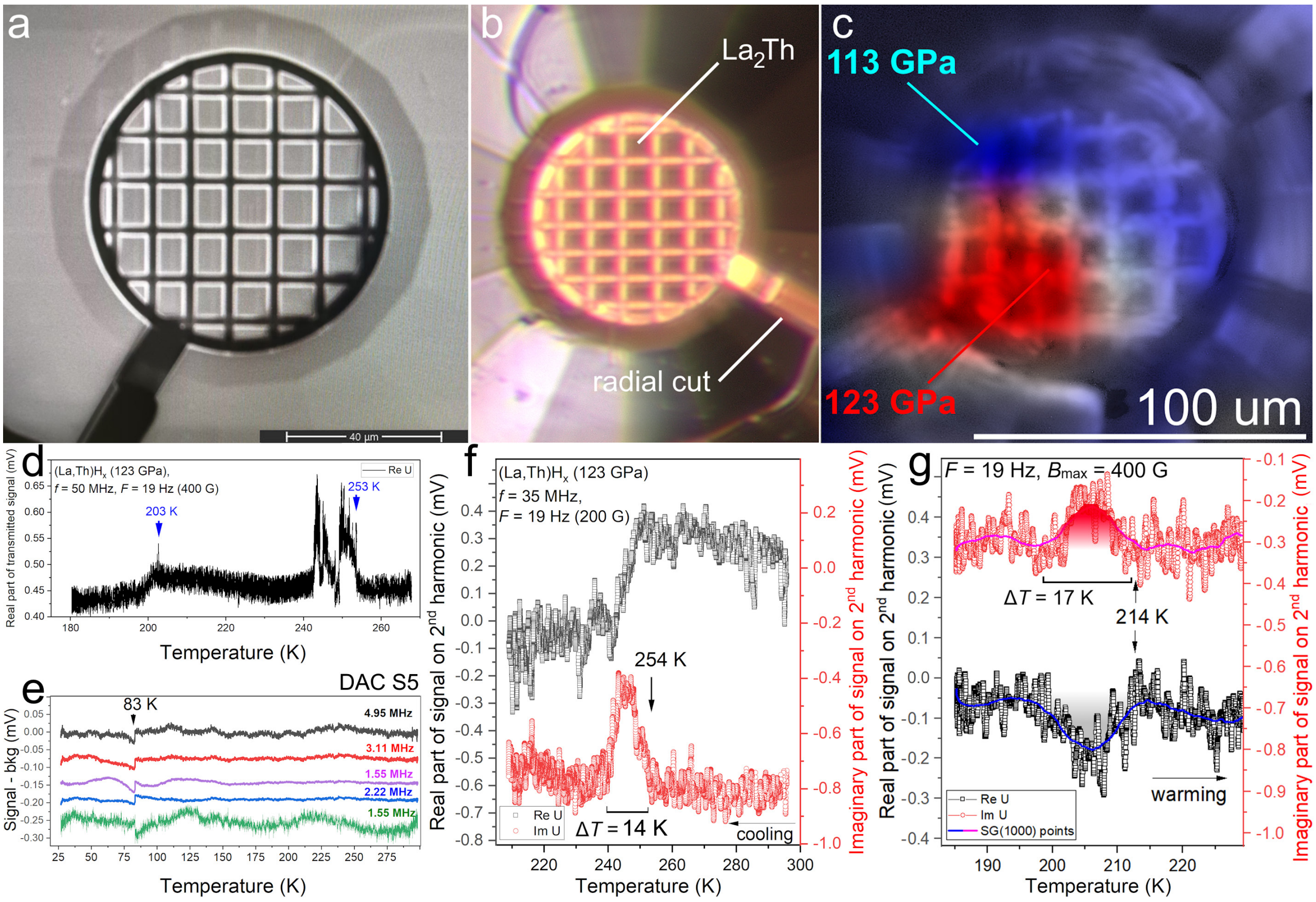

**Figure 4.** Probing superconductivity in La-Th hydrides under pressure of 113-123 GPa using RF transmission/reflection techniques. (a) Scanning electron microscopy image of the Ga ion-etched $La_2Th$ sample and the Lenz lens. (b) Optical micrograph of the sample after closure of the DAC and pressurization to 125 GPa, prior to laser heating. (c) Colored image of the multi-pixel $(La,Th)H_x$ matrix after laser heating of the central pixels, showing a pressure gradient (red marks higher pressure, maximum is about 123 GPa; blue corresponds to lower pressures, minimum is about 113 GPa). This pressure gradient may be the reason for the difference in the chemical composition of the synthesized hydrides in the blue and red zones of the sample. (d) The real part of the RF transmission signal of the DAC SM sample at 123 GPa as a function of temperature in the cooling mode. The carrier frequency is $f$ = 50 MHz. The contributions of the two superconducting phases (~203 K and ~253 K) are visible simultaneously. (e) RF reflection anomaly in DAC S5 at 83 K and 81 GPa measured simultaneously at four carrier frequencies (1.55, 2.22, 3.11, 4.95 MHz) in cooling mode. For clarity, the polynomial background has been subtracted ("Signal - bkg"). (f) Second harmonic signal ($2F$ = 38 Hz) of the low-frequency modulation field ($B_{max} \approx$ 200 G) in the RF transmission experiment with DAC SM at the carrier frequency of 35 MHz. At this frequency, the second harmonic signal for a high-temperature RF anomaly is well resolved. (g) A similar signal at the 2nd harmonic $2F$ = 38 Hz in a stronger field ($B_{max} \approx$ 400 G) observed in the 195-214 K region, typical for superconductivity in $(La,Th)H_{9-10}$.

Initially, the DAC SM was planned as a prototype of a multipixel superconducting RF detector (Figure 4). The matrix was fabricated using controlled ion etching (Ga FIB) of a sputtered film of $La_2Th$ sample (Figure 5a). Lenz lenses in their usual geometry[56] and 32 square pixels with a side of about 5-7 µm and 1-2 µm thickness, were cut from $La_2Th$ and Ta/Au films. The central part of the optical matrix, after compression with ammonia borane to 125 GPa, was heated pixel-by-pixel using a 1.06 μm IR laser to form lanthanum-thorium superhydrides (Figures 4b-c). The DAC SM had two Lenz lenses and 2 microcoils and was examined in the RF transmittance mode, whereas the DAC S5 had only one lens and inductive coil, and was examined only in RF reflection mode (Figure 4e, Supporting Figure S27).

Despite the presence of a number of special gaps between pixels in the (La,Th)$H_x$ matrix (Figure 4c), the sample demonstrates two the most pronounced anomalies in RF transmission both in the high-frequency channel (e.g., at 50 MHz, Figure 4d) and at the 2$^{nd}$ harmonic of the low-frequency modulating field (19 Hz, $B_{\max}$ = 200-400 Gauss, Figures 4f-g). The first RF signal at ≈ 203 K begins to appear already at 214 K (Figure 4g, Supporting Figure S18), corresponds to resistive measurement data (Figure 5a), and should be attributed to (La,Th)$H_{9\text{-}10}$. Meanwhile, a higher-temperature signal below 253-254 K in DACs S1-S3 with electrodes is observed only as an electrical resistance anomaly or the appearance of the very first signs of superconductivity in *dR/dT* (Supporting Figures S22b, S25b). As can be seen, this weak signal in *R(T)*, can be detected very reliably in the RF method and should be preliminarily associated with the hexagonal or *Fmmm* (La,Th)$H_{16}$ phase, detected by the XRD (Supporting Figures S11, S13). It would be appropriate to say that theoretical calculations for the structurally similar class of $XH_{18}$ polyhydrides (where X = La, Th, …) lead to very high $T_c$ = 271-306 K [57].

The width of the 2$^{nd}$ harmonic signal ($\Delta T \approx$ 14-17 K, Figures 4f-g) in order of magnitude is close to the width of the superconducting transition in inhomogeneous (La,Th)$H_{9\text{-}10}$ samples observed in transport measurements (Figures 3a, 5a), and can be used for rough penetration field assessment $B_p(0) \approx B_{\max} \cdot (T_c/2\cdot\Delta T)$ ≈ 0.18 – 0.25 Tesla. This estimate is based on the fact that only in the vicinity of $T_c$, when all critical fields are small, a weak modulating magnetic field can still penetrate the sample and suppress superconductivity in it, affecting the passage of the radio-frequency waves. The presence of a signal at the second harmonic of the modulating field also means that the signal in the high-frequency channel depends on the external magnetic field, which can be checked directly in a suitable frequency range. Indeed, by cooling the sample in a strong constant magnetic field from 0 to 4 T at frequencies of 10 and 75 MHz, we found a shift of the RF transition to the region of lower temperatures (Supporting Figure S26).

Finally, the DAC S5 sample, containing cubic and hexagonal hydrides of composition $XH_{5.75\text{-}6.25}$, was studied at 81 GPa using radio-frequency reflectance in the frequency range from 1.55 to 4.95 MHz (Supporting Figure S27). Compared to $(La,Ce)_4H_{23}$ [26], where $T_c \approx$ 68 K is achieved at 67 GPa, the replacement of cerium with thorium leads to the expected increase in $T_c$ to 83 K at 81 GPa, which is detected by a single anomaly in radio-frequency reflectance at all studied frequencies (Figure 4e, Supporting Figure S27).

## 3. Discussion

First, we should discuss deviations in the volumes of the La-Th superhydride unit cells from the ab initio calculations (Figure 5d). For the well-known phases *Fm*$\bar{3}$*m*-$XH_{10}$ and *P*$6_3$*/mmc*-$XH_9$, this deviation reaches approximately 0.5-0.75 Å$^3$/(La,Th) and corresponds to about 0.5 additional H atom. The deviation for *I*4/*mmm*-$XH_{4+x}$ is similar (~0.5 Å$^3$), but has the opposite sign. How can this be explained? First, we do not account here for anharmonic effects in the H-sublattice, which typically increase the unit cell volume of polyhydrides by 0.5–0.7 Å$^3$/heavy atom. This is most important for phases XH9 and XH10 and will likely resolve any discrepancies with the calculations. The second factor is the solid-solution nature of the heavy-atom sublattice, which was not taken into account in our simple DFT calculations of the equation of state (Figure 5d). Instead, we used Th-substituted, fully ordered analogs of known lanthanum polyhydrides with a substitution concentration of 25–50% (from LaTh to $La_3Th$). Despite some difference in atomic volumes between La and Th, the effect of the Th-substitution on the unit cell volume is not very large (typically < 0.5 Å$^3$/heavy atom). Finally, the tendency of polyhydrides toward non-stoichiometric compositions is well known; for example, the composition of *Fm*$\bar{3}$*m*-$LaH_{10}$ usually varies within the range of $LaH_{9\text{–}12}$ [17,58]. The same applies to the $X_3H_{11}$ phase, whose non-stoichiometric nature was noted earlier [20].

In this study, we synthesized La-Th superhydrides that are stable at pressures of approximately 1 Mbar and superconducting at temperatures of 190-232 K and even higher in some phases. The synthesis pressure can be reduced to 70-80 GPa, with a corresponding decrease in $T_c$ to 83 K. Moreover, the initial $La_2Th$ thin films are stable in air for several hours, unlike films of La, Ce, and La-Ce alloys, so all operations with La-Th can be performed outside a glovebox. Despite the success, this experiment raises the question of the maximum $T_c$ achievable in superhydrides at a fixed pressure $P$.

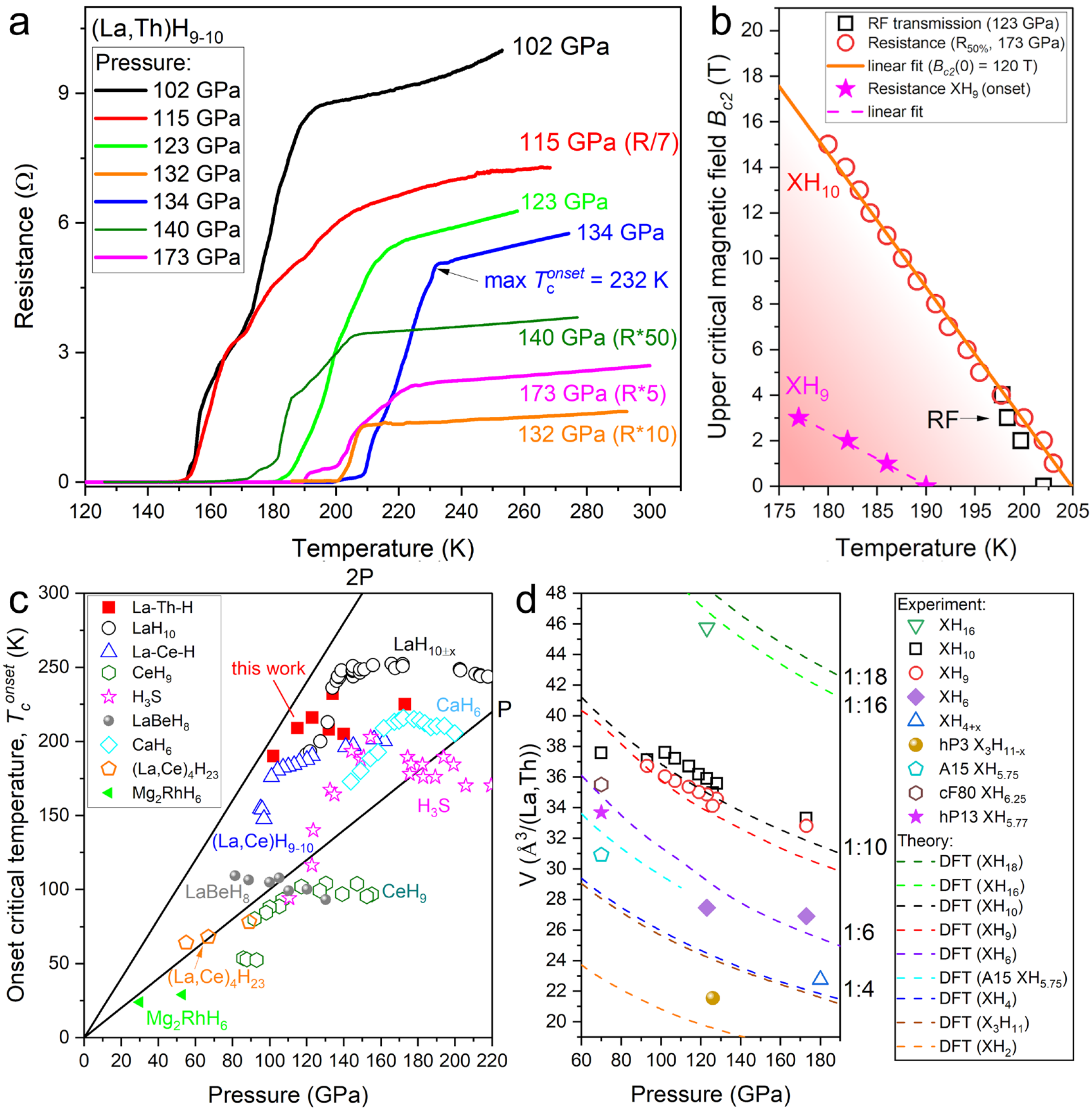

**Figure 5.** Superconductivity and unit cell volume of La-Th hydrides at high pressure. (a) Temperature dependence of the electrical resistance of several different DACs with $(La,Th)H_{9\text{-}10}$ at pressures from 102 to 173 GPa. Curves are rescaled for clarity. The maximum onset critical temperature in resistance measurements, $T_c$(onset) = 232 K was observed at 134 GPa in DAC S1. (b) Beginning of the superconducting phase diagram of $(La,Th)H_{9\text{-}10}$ constructed from transport and RF transmission measurements. The extrapolated upper critical magnetic field ($B_{c2}$) as a function of temperature is shown, along with a linear fit extrapolated to $B_{c2}(0) \approx 120$ T. (c) Critical temperatures $T_c$(onset) versus pressure for La–Th–H system (red squares, this work) compared with $LaH_{10}$ [2,7], $(La,Ce)H_{9\text{-}10}$ [8,9], $CeH_9$ [10], $H_3S$ [1], $LaBeH_8$ [59], $CaH_6$ [60,61], $(La,Ce)_4H_{23}$ [26] and $Mg_2RhH_6$ [62]. The straight lines show $T_c = P$ and $T_c = 2P$ (units are K and GPa). (d) The pressure dependence of the unit-cell volume per X = (La,Th) for the experimentally observed La-Th hydrides: $XH_{16-x}$, $XH_{10}$, $XH_9$, $XH_6$, $XH_{4+x}$, *hP*3 $X_3H_{11-x}$, A15 $X_4H_{23}$, *cF*80 $X_4H_{25}$ and *hP*13 $XH_{5.77}$ (Supporting Table S2) compared to approximate DFT equations of state (VASP code, dashed lines). The labels on the right (1:4 to 1:18) indicate the X:H ratio of the corresponding phases.

Despite the significant amount of accumulated experimental material, it is still not possible to cross the line *max* $T_c$ [in K] = 2$P$ [in GPa] (Figure 5c). Moreover, even if the most recent results on La-Sc-H system, $LaSc_2H_{24}$, and $LaH_{12}$, where $T_c$ may reach 270-300 K [48,51,63], are correct, the pressure required to achieve the superconducting state is still higher than 150-250 GPa. Thus, the upper limit of *max* $T_c = 2P$ is not reached.

The physical sense of the empirical rule *max* $T_c \approx 2P$ is simple. Considering that the critical temperature for Bardeen-Cooper-Schrieffer-Migdal-Eliashberg (BCS-ME) superconductors depends linearly on some average phonon energy [64] (Debye frequency $\theta_D$ or the logarithmically averaged frequency $\omega_{log}$), and this phonon frequency depends on the interatomic distances, we can write

$$T_c = a'\theta_D \propto aP^{\alpha} + b, \qquad (3)$$

where we have taken into account that the interatomic distances depend on a volume of a unit cell *V(P)*, which is compressed under pressure following an equation of state. Neglecting the difference between the exponent $\alpha$ and 1, we obtain $T_c \propto P$. From experimental data it follows that for the best hydride superconductors have the coefficient $a < 2$ [K/GPa], and the highest critical temperature at ambient pressure gives $b = 40$ K, and belongs to well-known BCS-ME superconductor $MgB_2$ [65].

It is important that the main conclusion can be made even by setting $b = 0$ in the eq. (3): from the experimental data it follows that superhydrides cannot be stabilized at pressures below 70 GPa while maintaining a critical temperature well above 140 K and, thus, will be inferior to cuprates and nickelate superconductors at pressures when diamond anvils can be reversible decompressed. Indeed, DFT predictions of high-$T_c$ superconductivity at pressures below 50 GPa lead to $T_c \sim 100$ K [66]. However, most of them likely have no chance of experimental confirmation.

Eq. 3 is quite simple, although it describes the experimental trend. A bit more accurate formula can be easily obtained if we use the Grüneisen parameter $\gamma = -\,\partial\ln(\omega)/\partial\ln(V)$ and the power-law approximation of the Birch-Murnaghan [67,68] equation of state:

$$V \approx V_0\left(1+\frac{B_0'}{B_0}P\right)^{-\frac{1}{B_0'}} \to \omega(P) \approx \omega(0)\left(1+\frac{B_0'}{B_0}P\right)^{\frac{\gamma}{B_0'}} \to T_c = T_0(1+a''P)^d, \qquad (4)$$

where $B_0$ is the bulk modulus of the material at ambient conditions, and $B_0'$ is the pressure derivative of the bulk modulus at zero pressure, $d = \frac{\gamma}{B_0'}$. Typical values of $B_0' = 4.5 - 6$ [69]. For metals and lower hydrides of metals $B_0 \approx 50 - 150$ GPa [68,70,71], and $\gamma = 1 - 2$ [68,70,71]. In the problem of reducing the stabilization pressure $P$ considerably lower than $B_0$, the term $\propto P$ in the eq. (4) turns out to be less than unity, which, when using the Taylor expansion, brings us back to the eq. (3).

## Conclusions

In this work, we show that partial replacement of La in $LaH_{10}$ with thorium yields a new family of lanthanum-thorium superhydrides with a mixed-metal sublattice. The cubic *fcc* (La, Th)$H_{10}$ and hexagonal *hcp* (La, Th)$H_9$ phases can be synthesized and remain stable at pressures near 1 Mbar. $T_c$ of cubic $XH_{10}$ reaches 232 K at 134 GPa, which is higher than in the Ce-containing analogues at similar pressures. The decahydride persists on decompression to at least 93-95 GPa and the nonahydride to about 90 GPa. Superconductivity was established by two independent approaches, charge transport in magnetic fields and contactless radio-frequency detection, and both give a consistent picture of the superconducting transition and its field dependence. Pulsed current measurements up to 0.38 A in (La,Th)$H_9$ at 173 GPa provide estimates of s-wave single superconducting gap $\Delta(0) = 37 \pm 4.5$ meV, and $2\Delta(0)/kT_c = 4.9 \pm 0.6$, in accordance with theoretical calculations and expectations.

The chemistry of La-Th polyhydrides is rich. Beyond the main phases, we observed formation of hexagonal $(La,Th)_3H_{11-x}$, nanocrystalline *I*4/*mmm*-(La,Th)$H_4$ (5-6 nm grains), *hP*13 $(La,Th)_{13}H_{75}$, cubic *cF*80 $(La,Th)_4H_{25}$ and A15 $(La,Th)_4H_{23}$ stable at 70 GPa, and possibly (La,Th)$H_{16}$, responsible for a strong signal in RF transmission near 253 K. The $XH_{5.75\text{-}6.25}$ phases show $T_c \approx 83$ K at 81 GPa. Some non-stoichiometry is observed for many compositions within the range of ±1 H. In addition to all of the above, sputtered films of $La_2Th$ and $La_3Th$ alloys are air-stable for hours, so handling them does not require a glovebox, which makes them very convenient for the synthesis of superhydrides.

**Acknowledgements**

D.V.S. and D.Z. would like to express their gratitude to the National Natural Science Foundation of China (NSFC), Research Fund for International Excellent Young Scientists, grant number W2632008, for the financial support. D.V.S., D.Z., V.V.S., and I.A.T. thank the NSFC International Collaboration Fund for Creative Research Teams, grant No. W2641002, for financial support. This work was supported by the National Key Research and Development Program of China (grant 2023YFA1608900, subproject 2023YFA1608903). V.V.S. acknowledges financial support from the Shanghai Science and Technology Committee, China (No. 22JC1410300) and Shanghai Key Laboratory of Materials Frontier Research in Extreme Environments (MFree), China (No. 22dz2260800). I.A.T., A.V.S., A.T.D., A.S.U., V.M.P acknowledge the financial support of RSF grant 26-12-00447. D.S. and D.Z. are grateful for financial support from the EMFL-ISABEL project. We acknowledge support from the Würzburg-Dresden Cluster of Excellence on Complexity, Topology and Dynamics in Quantum Matter–ctd.qmat (EXC 2147, Project No. 390858490), as well as the support of the HLD at HZDR, member of the European Magnetic Field Laboratory (EMFL). The high-pressure experiments were carried out within the state assignment of the Institute for Nuclear Research of the Russian Academy of Sciences. The authors express their gratitude to the Collective Facility Center of the Institute of Physical Chemistry and Electrochemistry of the Russian Academy of Sciences for assistance in the synthesis of La-Th alloys. This research used facilities of the Brazilian Synchrotron Light Laboratory (LNLS), part of the Brazilian Center for Research in Energy and Materials (CNPEM), a private non-profit organization under the supervision of the Brazilian Ministry for Science, Technology, and Innovations (MCTI). All X-ray diffraction experiments were performed at the EMA beamline of the Sirius synchrotron light source (Campinas, Brazil) under proposal EMA-20260950. We are particularly grateful to Rodolfo T. Souza and Joel A. V. Mendonça for their assistance during the experiments. All operations with thorium metal and its alloys were carried out at the Institute of Physical Chemistry and Electrochemistry of the Russian Academy of Sciences (Moscow).

**Author contributions**

D.V.S.: conceptualization; methodology; investigation (DAC preparation, XRD, transport measurements, radio-frequency studies); formal analysis (XRD refinements, $T_c(P)$, empirical $T_c$-$P$ analysis); writing - original draft; supervision. M.A.V., M.I.V., I.M.N.: resources (synthesis and characterization of La-Th alloys, handling of thorium and alloys); investigation. A.V.S., A.T.D., A.S.U.: investigation (pulsed-current measurements); formal analysis ($J_c(T)$, $\Delta(0)$, $\lambda_L(0)$); methodology (pulsed current source).: investigation (electrical and pulsed-current measurements); V.M.P.: methodology; supervision (transport measurements); writing - review and editing. T.H.: investigation (steady-field up to 16 T at HLD); resources. D.Z.: investigation (DACs preparation, synchrotron XRD studies); funding acquisition. V.V.S.: conceptualization; methodology (RF technique); supervision; funding acquisition; writing - review and editing. I.A.T.: investigation (DACs preparation, transport measurements); supervision; funding acquisition; writing - review and editing. All authors discussed the results and approved the final manuscript.

**Data availability**

The authors declare that the main data supporting findings of this study are contained within the paper and the Supporting Information. Data are available from the corresponding author upon reasonable request.

**Competing interests**

The authors declare no conflicts of interest.

# SUPPORTING INFORMATION

## Stabilization of $LaH_{10}$ with Thorium:

## Superconductivity in $(La,Th)H_{10}$ up to 232 K near 1 Mbar

Dmitrii V. Semenok[1,†,*], Mikhail A. Volkov[2,†], Andrey V. Sadakov[3,†], Alexander T. Daniyarkhodzhaev[3], Alexey S. Usoltsev[3], Vladimir M. Pudalov[3], Mikhail I. Volgin[2], Iurii M. Nevolin[2], Toni Helm [4], Di Zhou[1], Viktor V. Struzhkin [5,*], and Ivan A. Troyan [6,*]

[1]*Center for High Pressure Science and Technology Advanced Research (HPSTAR), Beijing*

[2] *Institute of Physical Chemistry and Electrochemistry of the Russian Academy of Sciences, 119071, Moscow, Leninsky Prospekt, 31/4, Russia*

[3] *V. L. Ginzburg Center for High-Temperature Superconductivity and Quantum Materials, Moscow, 119991 Russia*

[4] *Dresden High Magnetic Field Laboratory (HLD-EMFL), Helmholtz-Zentrum Dresden-Rossendorf (HZDR), 01328 Dresden, Germany*

[5] *Center for High Pressure Science and Technology Advanced Research (HPSTAR), Shanghai*

[6] *Shanghai Key Laboratory of Material Frontiers Research in Extreme Environments (MFree), Shanghai Advanced Research in Physical Sciences (SHARPS), Pudong, Shanghai 201203, China*

*Corresponding authors, emails: dmitrii.semenok@hpstar.ac.cn (Dmitrii V. Semenok), viktor.struzhkin@hpstar.ac.cn (Viktor V. Struzhkin), and itrojan@sharps.ac.cn (Ivan A. Troyan)

[†]These authors contributed equally to this work

# Contents

## 1. Methods

*Experiment*

Several diamond anvil cells (DACs) were prepared for synthesis and structural/transport characterization of La–Th hydrides under high pressure (Table S1).

DACs S0 and S5 was prepared for low-pressure (100-140 GPa and below) X-ray diffraction (XRD) studies. DAC S0 utilized a CuTi cell body, a steel/$CaF_2$/epoxy insulating gasket, 50 um culet diamond anvils, and contained no electrodes. XRD analysis upon decompression from 128 GPa to 93 GPa revealed the presence of *fcc*-$XH_{10}$, *hcp*-$XH_9$, and side phases. DAC S5 was a NiCrAl BX-90 mini cell with 75 µm diamond anvil culets, equipped with one Ta/Pt Lenz lens, pick-up microcoil with 30 turns, and $TaW_{10\%}$ oxidized gasket.

DACs S1 and S2 were assembled using BeCu cell bodies (d = 25 mm) and 50 um culet anvils to investigate the superconducting properties of La–Th hydrides at lower pressure range (100–140 GPa) using Ta/Au electrical leads. Insulating gaskets were composed of W/$CaF_2$/epoxy for S1 and steel/$CaF_2$/epoxy for S2.

DAC S3 (BeCu cell, steel/$CaF_2$/epoxy gasket, Ta/Au electrodes, 50 um culets) was designed to explore the high-pressure region up to 173 GPa. At 173 GPa, XRD confirmed the formation of *fcc*-$XH_{10}$ and *hcp*-$XH_9$ phases.

DAC SM (matrix) was prepared for radio-frequency (RF) studies and to evaluate the effect of excess ammonia borane at the pixels' edge. It featured larger diamond culets (100 um diameter) to provide the largest sample area among all DACs. The cell utilized a BeCu body, a W/Ta/$Ta_2O_5$ insulating gasket, Ta/Au/$La_2Th$ Lenz lenses, and a Ga-treated $La_2Th$ starting material (6-13 at.% residual Ga content introduced via focused ion beam milling). The synthesized phase was measured at the pressure of 113–123 GPa.

Pulsed field DAC (NiCrAl cell body, W/$CaF_2$/epoxy gasket, Ta/Au electrodes, 50 um culets, tested at 137–132 GPa) was prepared using $La_3Th$ starting alloy for high-magnetic-field measurements in pulsed fields. However, due to the electrode system damage, pulsed-field electrical transport measurements could not be completed.

**Table S1.** Parameters of samples, high-pressure DACs, and diamond anvils used in this study.

| | **Starting material** | **DAC's material** | **Insulating gasket** | **Electrodes** | **Culet diameter, µm** | **Pressure (± 3 GPa)** | **Composition according to XRD** |
|---|---|---|---|---|---|---|---|
| **DAC S0** | $La_2Th$/AB | CuTi | Steel/$CaF_2$/epoxy | no | 50 | decompression from 128 GPa to 93 GPa | *fcc*-$XH_{10}$ + *hcp*-$XH_9$ + side phases |
| **DAC S1** | $La_2Th$/AB | BeCu | W/$CaF_2$/epoxy | Ta/Au | 50 | 102-134 GPa | *fcc*-$XH_{10}$ + *hcp*-$XH_9$ + *hP*3-$X_3H_{11}$ |
| **DAC S2** | $La_2Th$/AB | BeCu | Steel/$CaF_2$/epoxy | Ta/Au | 50 | 123, 140 GPa | *fcc*-$XH_{10}$ + *hcp*-$XH_9$ |
| **DAC S3** | $La_2Th$/AB | BeCu | Steel/$CaF_2$/epoxy | Ta/Au | 50 | 173 GPa | *fcc*-$XH_{10}$ + *hcp*-$XH_9$ + $XH_6$ |
| **DAC SM (matrix)** | $La_2Th$(Ga) /AB* | BeCu | W/Ta/$Ta_2O_5$ | Lenz lenses: Ta/Au/$La_2Th$ | 100 | 113-123 GPa | *fcc*-$XH_{10}$, $XH_6$, $XH_{16}$ |
| **Pulsed field DAC** | $La_3Th$/AB | NiCrAl | W/$CaF_2$/epoxy | Ta/Au | 50 | 137 → 132 GPa | *fcc*-$XH_{10}$ + *hcp*-$XH_9$ + *hcp*-$XH_6$ |
| **DAC S4** | $La_2Th$/AB | BeCu | Steel/$CaF_2$/epoxy | Ta/Au | 50 | 174 → 180 GPa | *bct*-$XH_4$ |
| **DAC S5** | $La_5Th$/AB | NiCrAl | $TaW_{10\%}$/$Ta_2O_5$ | Lenz lens: Ta/Pt | 75 | 70 → 81 GPa | *cF80, hP13, A15, fcc*-$XH_{10-x}$ |

* Ga concentration after Ga FIB treatment of the sputtered $La_2Th$ surface layer is about 6-13 at%.

Considering the multiple synchrotron experiments, DACs transportation, measurements on different Raman systems, pressure gradient from the center to the cell boundaries, and multiple cooling and heating cycles, the DACs pressure was not constant, but varied within ±3 GPa, which was reflected in different pressures for the same DACs in different experiments.

To enhance our ability to measure critical current beyond the capability of the Keithley 6221, we engineered our custom current booster, or, in other words, a current-controlled current source (CCCS). This CCCS allowed us to source current into the sample up to 0.6 A and perform voltage sweeps up to 60 V, while preserving the Keithley 6221's setup speed and resolution. The design was implemented with precautions that ensure the Keithley 2182a can operate within its high-resolution range (Figure S1). The current pulse duration was typically 10-50 microseconds, and the interval between pulses was 10-50 milliseconds. Since La-Th hydrides does not fill the entire volume of the chamber, the actual cross-sectional area of the current bridge is less than 1.5×7 $\mu m^2$. Thus, in our V-I measurements, we are coming to a lower bound for the critical current density.

The current source operates in such a way that during idle time, the sample is held at the supply potential. When a measurement pulse is applied, the sample's potential shifts abruptly, which introduces a small artifact into the measurement circuit. Furthermore, the current-carrying wires and sample contacts collectively possess a resistance on the order of 100 Ohms. At currents of approximately 100 mA, this results in a common-mode voltage ($V_{CM}$) of up to 10 V. Typically, modern integrated amplifiers have a common-mode rejection ratio (CMRR) of around $10^6$ (120 dB). Thus, 10V/$10^6$ results in an offset of ~10 μV. The monotonic offset on the order of several μV observed with increasing current is attributed precisely to this effect and does not affect the accuracy of the critical current determination.

There are several techniques to probe the order parameter in superconductors under pressure. Some of them are based on tunnel effects[1], while others are based on the investigation of the temperature dependence of the superfluid density. The latter proved to be a reliable tool to determine not only the quantity and values of energy gaps, but also the symmetry of the order parameter – whether it has s-wave, d-wave, or even more complex structure[2]. This usually requires either measuring directly the London penetration depth $\lambda(T)$, or the lower critical field, or measuring the self-field critical current density $J_c(T)$. Given all the limits and difficulties of measurements in DAC under high pressure, we were only able to measure the self-field critical current.

Talantsev *et al* suggested that several superconducting parameters such as the energy gap and the London penetration depth can be derived from the temperature dependence of the self-field critical current density. The parameters obtain by this method for conventional superconductors[3], for copper oxide high temperature superconductors[2] and for iron based high temperature superconductors[4] are in a good agreement with available data.

For a type II superconductor with the rectangular cross section critical current density is given by[5]

$$J_c(T) = \frac{\Phi_0(ln(\kappa)+0.5)}{4\pi\mu_0\lambda^3(T)}\left(\frac{\lambda(T)}{a}tanh\left(\frac{a}{\lambda(T)}\right)+\frac{\lambda(T)}{b}tanh\left(\frac{b}{\lambda(T)}\right)\right), \quad (S1)$$

where $\mu_0$ is vacuum permeability, $\Phi_0$ is the flux quantum, $2a$ and $2b$ are the width and the height of the sample, respectively, $\lambda(T)$ is London penetration depth and $\kappa$ is Ginzburg-Landau parameter. The latter has a weak temperature dependence and thus $ln(\kappa)$ can be thought as being effectively constant.

In the case of a cylindrical Fermi surface the temperature dependence of the normalized superfluid density is given by[5]

$$\rho_S(T) = 1 - \frac{1}{2\pi k_B T}\int_0^{2\pi} d\theta\, cos^2(\theta) \times \int_0^{\infty} cosh^{-2}\left(\frac{\sqrt{\varepsilon^2+\Delta^2(T,\theta)}}{2k_B T}\right) d\varepsilon, \quad (S2)$$

where $k_B$ is the Boltzmann constant and $\Delta(T,\theta)$ is an angle dependent gap function. For an isotropic Fermi surface the expression becomes

$$\rho_S(T) = \frac{\lambda^2(0)}{\lambda^2(T)} = 1 - \frac{1}{2k_B T}\int_0^{\infty} cosh^{-2}\left(\frac{\sqrt{\varepsilon^2 + \Delta^2(T)}}{2k_B T}\right) d\varepsilon. \quad (S3)$$

The temperature dependence of the superconducting gap can be approximated by the expression given by Gross *et al* [6]

$$\Delta(T) = \Delta(0) tanh\left(\frac{\pi k_B T_c}{\Delta(0)}\sqrt{a(T_c/T - 1)}\right) \quad (S4)$$

where $T_c$, $2\Delta(0)/k_B T_c$ and $a$ are fitting parameters. An angle dependent gap function $\Delta(T,\theta)$ can be represented $\Delta(T,\theta) = \Delta(T)\cdot g(\theta)$ with $\Delta(T)$ being the temperature dependence of the gap function and $g(\theta)$ the unity magnitude angular dependence. For an anisotropic s-wave case $g(\theta) = \big(1 + mcos(4\theta)\big)/(1+m)$ with $m$ being anisotropy parameter and for the d-wave pairing $g(\theta) = cos(2\theta)$.

The temperature dependence of the critical current density was obtained from current-voltage characteristics measured from 10 K up to $T_c$. Critical current was determined from the current voltage characteristic after offset subtraction by the current when voltage on the sample reaches 2 μV which is a widely used criterion in the literature[7]. The sample cross section for the calculation of the critical current density was determined from the geometrical parameters.

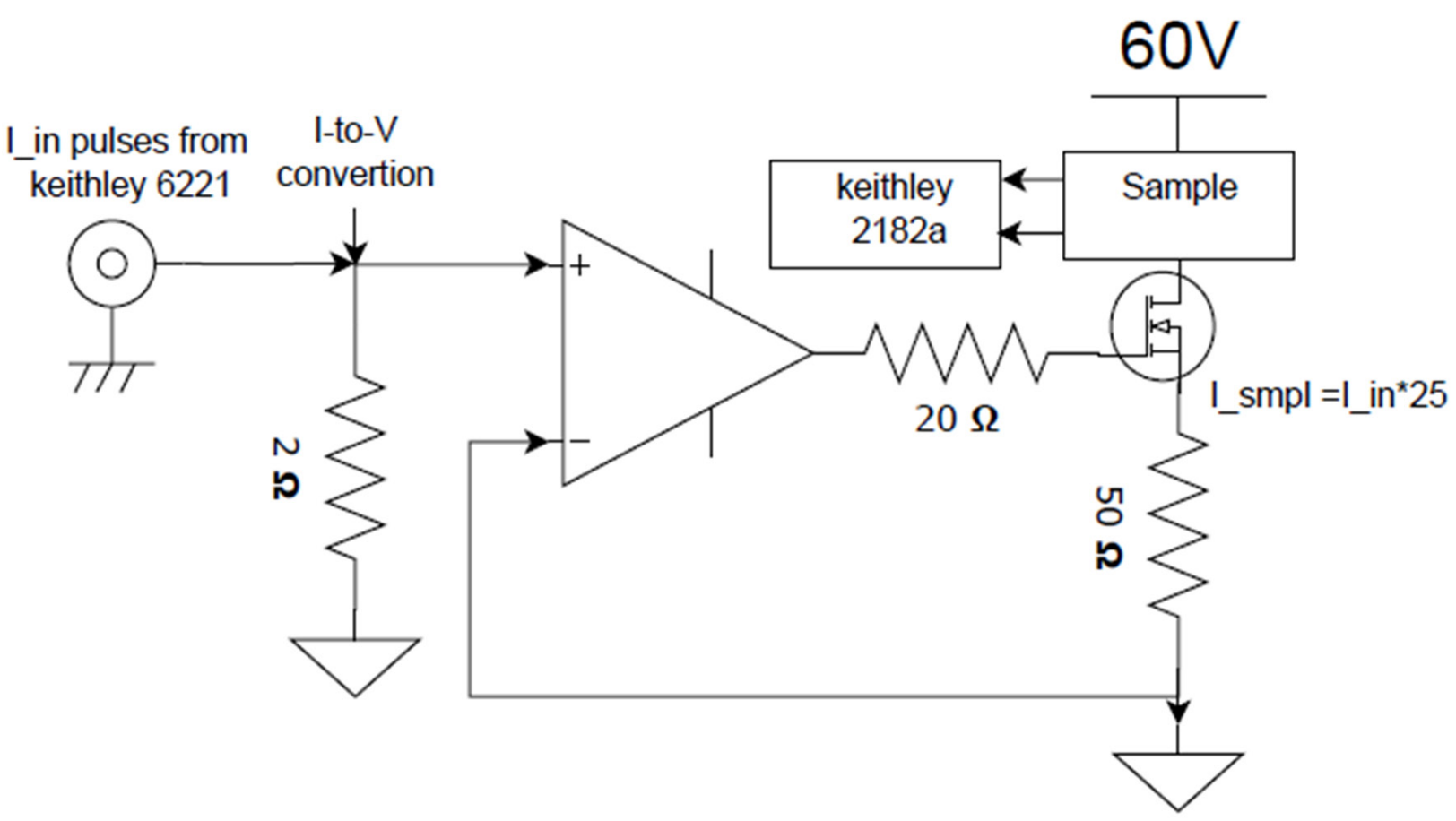

**Figure S1.** Scheme of measurements of voltage-current characteristics of the DAC S3 sample in the pulse mode using an amplifier of rectangular current pulses received from Keithley 6221.

*Theory*

Equation of state (EOS) calculations were performed using density functional theory (DFT) as implemented in the Vienna Ab initio Simulation Package (VASP) [8-10]. The projector augmented wave (PAW) method with the PBE exchange-correlation functional [11] was employed to describe the ion-electron interactions. A plane-wave basis set with a cutoff energy of 600 eV was used. The Brillouin zone was sampled using a reciprocal-space resolution of 0.25 Å$^{-1}$. To determine the equilibrium volume, the cell shape, volume, and internal atomic positions were fully relaxed using a conjugate-gradient algorithm (IBRION = 2, ISIF = 3), with the ionic relaxation continued until all forces converged. For the final static energy calculations, the tetrahedron method with Blöchl corrections (ISMEAR = -5) was used to obtain accurate total energies.

First-principles rough estimations of the electron–phonon coupling in $LaThH_{20}$, $LaThH_{18}$, $YThH_{18}$ at 180-200 GPa were performed using the Quantum ESPRESSO package[12-14]. The self-consistent field calculations employed a plane-wave kinetic-energy cutoff of 80 Ry and a 16 × 16 × 8 Monkhorst–Pack k-point mesh. Projector augmented-wave (PAW) pseudopotentials[15] were used. The thorium pseudopotential was generated using a scalar-relativistic calculation within the Perdew–Burke–Ernzerhof (PBE) generalized-gradient approximation and included the 6s, 7s, 6p, 6d, and 5f states in the valence configuration. Phonon calculations were performed on a 4 × 4 × 2 q-point grid. The electron–phonon coupling was evaluated using the tetrahedron method [16].

## 2. Elemental analysis

Ingots of lanthanum (grade LaM-1) and thorium (purified by zone melting) were drilled to produce shavings inside an argon-filled SPEKS GB (Russia) glove box. The lanthanum shavings were removed from the glove box and melted into a button using a Centorr 5SA electric arc furnace (US) in an atmosphere of high-purity argon with a zirconium getter prior melting. The resulting button was weighed; the required mass of thorium was then calculated, and a corresponding amount of thorium shavings was prepared inside the glove box. La–Th alloys were produced by melting the lanthanum button together with the thorium shavings in an electric arc furnace with the resulting combined button being re-melted five times. Mass loss during the melting process was approximately 1%.

The EDX spectra were acquired at 10 kV. For a $La_2Th$ film (density ≈7.8 g/cm$^3$) the Kanaya-Okayama electron range at this energy is about 0.7 µm [17] and the depth of X-ray generation is smaller, so the 1-2 µm thick film is only partly probed. Since the Ga K edge (10.37 keV) lies above the beam energy, Ga is detected only through its L lines (about 1.1 keV), which are strongly absorbed in the La-Th matrix and come from the top of the sample ~0.1 µm. Gallium is introduced by FIB milling as Ga+ implantation into a surface layer of 10-30 nm thickness for 30 keV ions in heavy metals [18]. The Ga concentration in Figure S2 is therefore an apparent near-surface value and overestimates its volume content. For an implanted layer of 10-30 nm on the top of the sample of 1-2 µm thickness, the layer occupies 1-3% of the volume, and the volume-averaged Ga content does not exceed a few at.%. The La:Th ratio from the same spectra is semi-quantitative (standardless analysis).

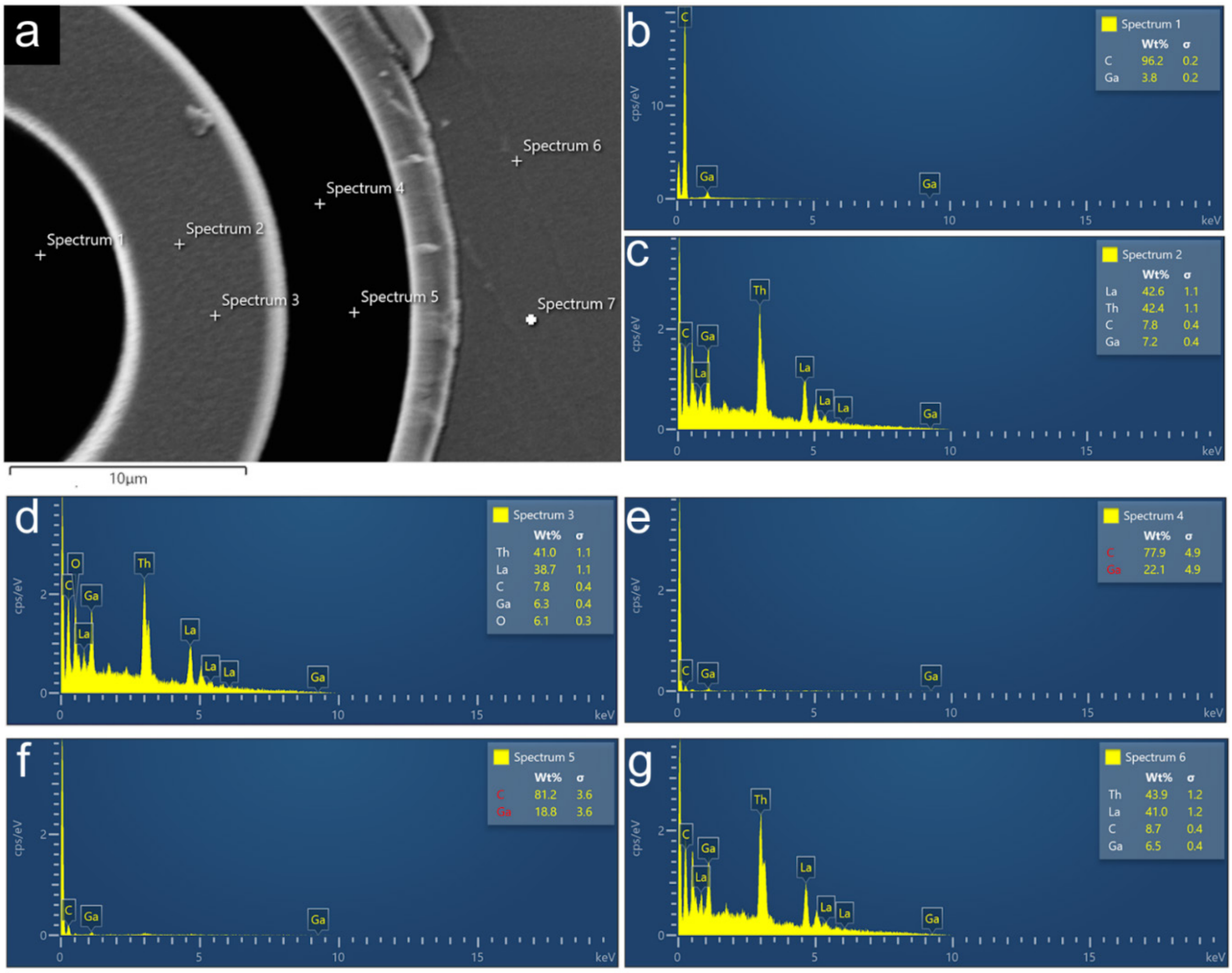

**Figure S2.** Example of the energy-dispersive X-ray spectroscopy (EDX) analysis of a $La_2Th$ thin film sputtered on a diamond anvil culet, with ring structures created using focused gallium ion beam (Ga+ FIB). (a) Scanning electron microscope image indicating seven specific measurement spots (Spectra 1-7) across the ring pattern and culet surface. (b-g) EDX spectra corresponding to selected regions: (b) Spectrum 1 on the carbon substrate; (c) Spectrum 2 on the alloy layer; (d) Spectrum 3 on the alloy layer; (e) Spectrum 4 on the outer surface; (f) Spectrum 5 on the surface edge; and (g) Spectrum 6 on the sample body. The analysis confirms a La to Th atomic ratio close to the $La_2Th$ stoichiometry, along with residual gallium contamination introduced during the FIB milling process.

### 3. Raman and optical spectroscopy

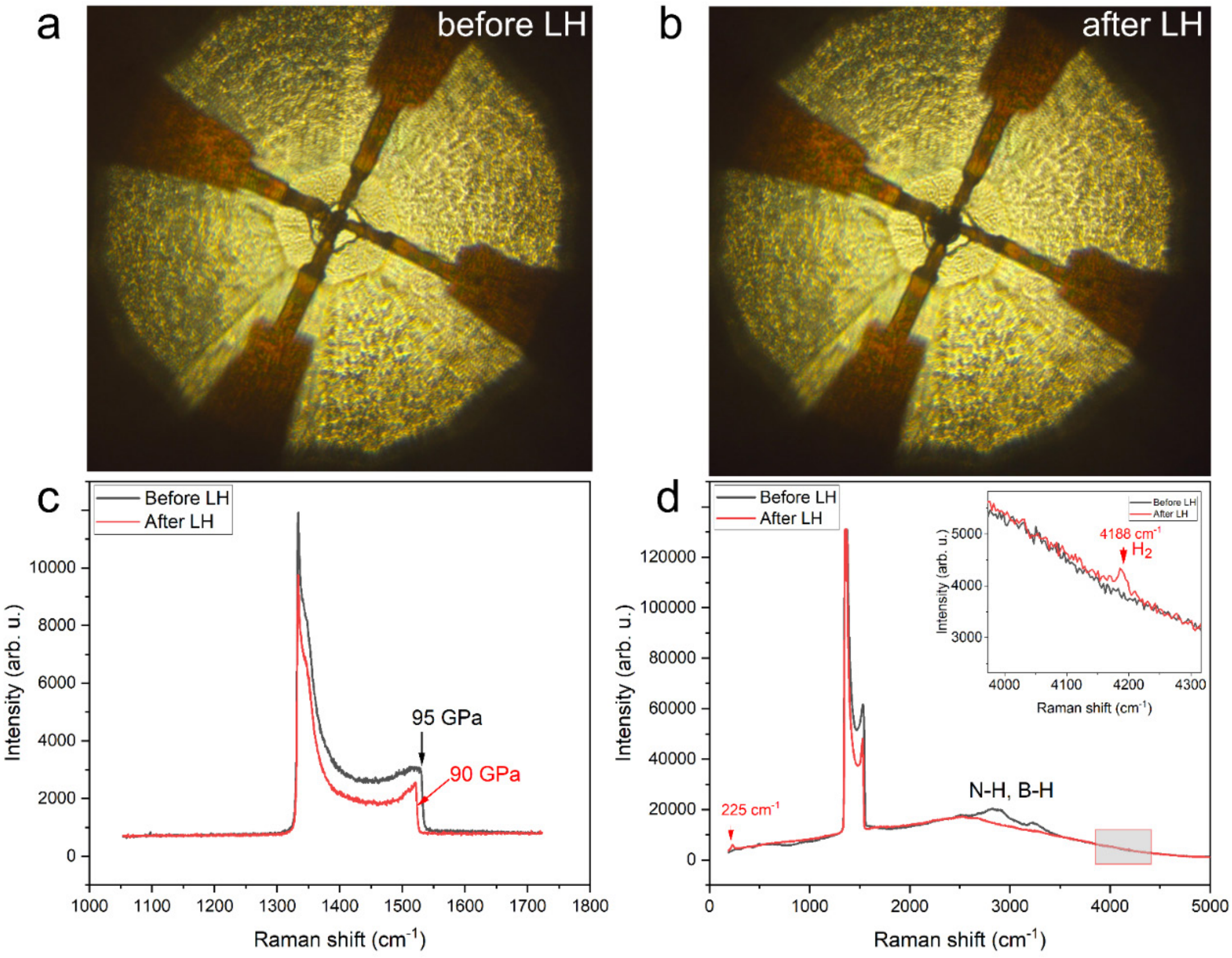

**Figure S3.** Photographs of La-Th/AB samples before and after laser heating at 90-95 GPa, and Raman spectroscopy of the synthesized samples. (a, b) Photographs of the culet before and after laser heating of a loaded $La_2Th$ particle. (c, d) Raman spectra of the center of high-pressure chamber before and after laser heating. A small molecular hydrogen signal is also detected (inset in panel "d"), whose position corresponds to a slightly higher pressure of about 100 GPa.

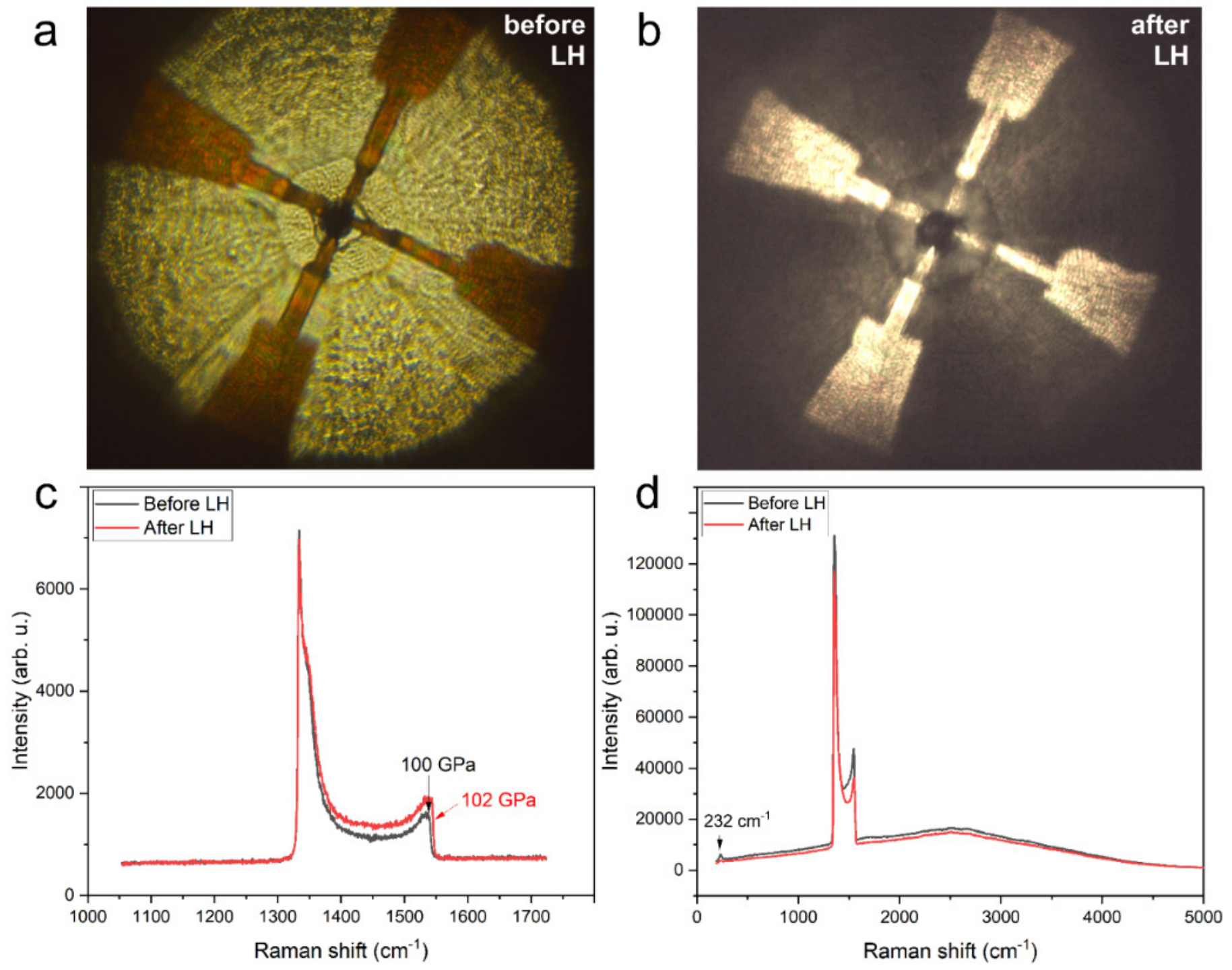

**Figure S4.** Photographs of La-Th/AB samples before and after laser heating at 100-102 GPa, and Raman spectroscopy of the synthesized samples. (a, b) Photographs of the culet before and after laser heating of a loaded $La_2Th$ particle. (c, d) Raman spectra of the center of high-pressure chamber before and after laser heating. A small signal at 232 $cm^{-1}$ is detected.

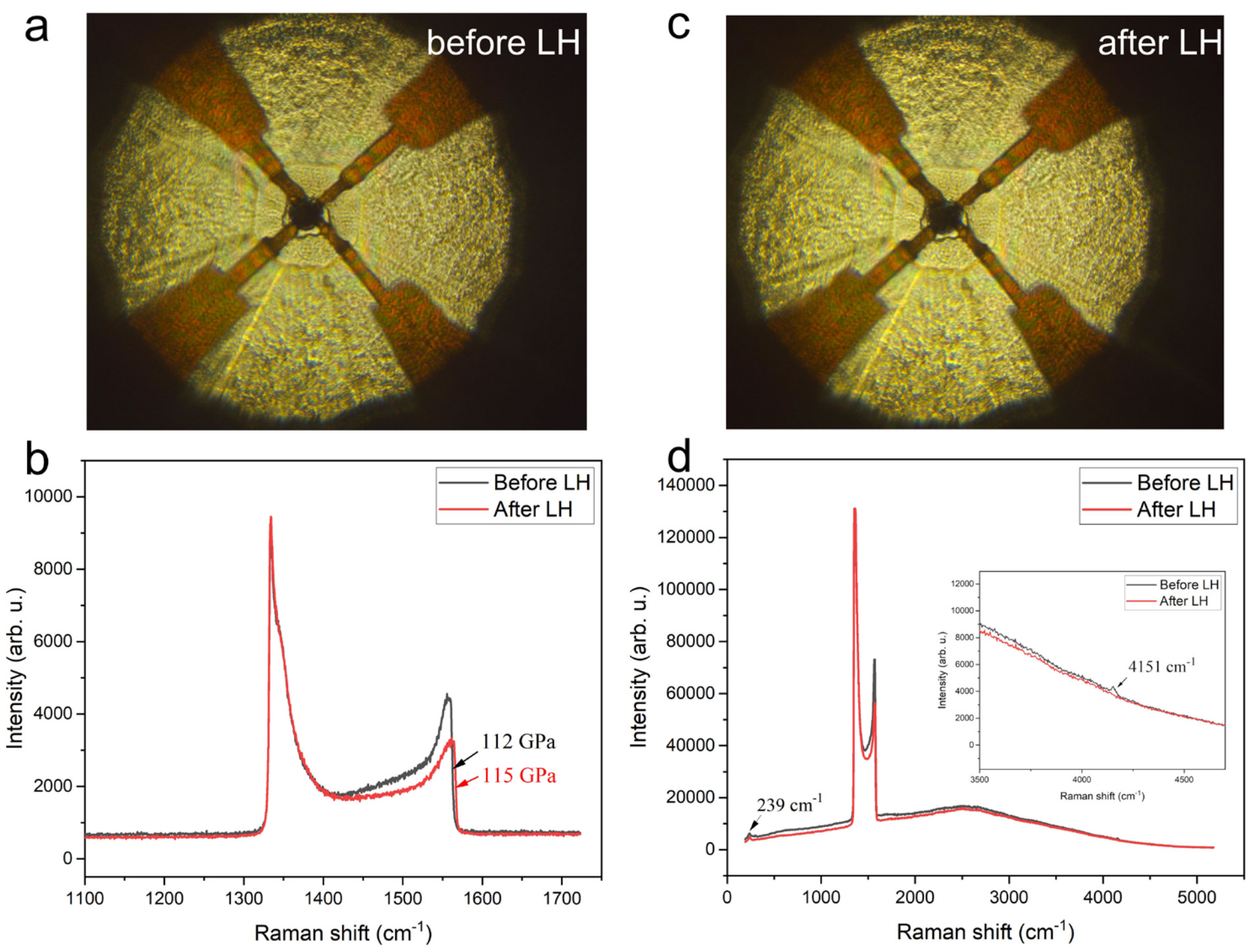

**Figure S5.** Photographs of La-Th hydride samples before and after additional (run 3) laser heating at 112-115 GPa, and Raman spectra of the synthesized samples. (a, c) Photographs of the samples before and after additional (run 3) laser heating of the loaded $La_2Th$ particle. (b, d) Raman spectra of the center of high-pressure cells before and after laser heating. A small signal is detected at 239 cm$^{-1}$. A signal of molecular hydrogen was also detected (insert in panel "d"), the position of which corresponds to a pressure of about 113 GPa. After additional laser heating, hydrogen is absorbed by the sample and disappears from the spectrum.

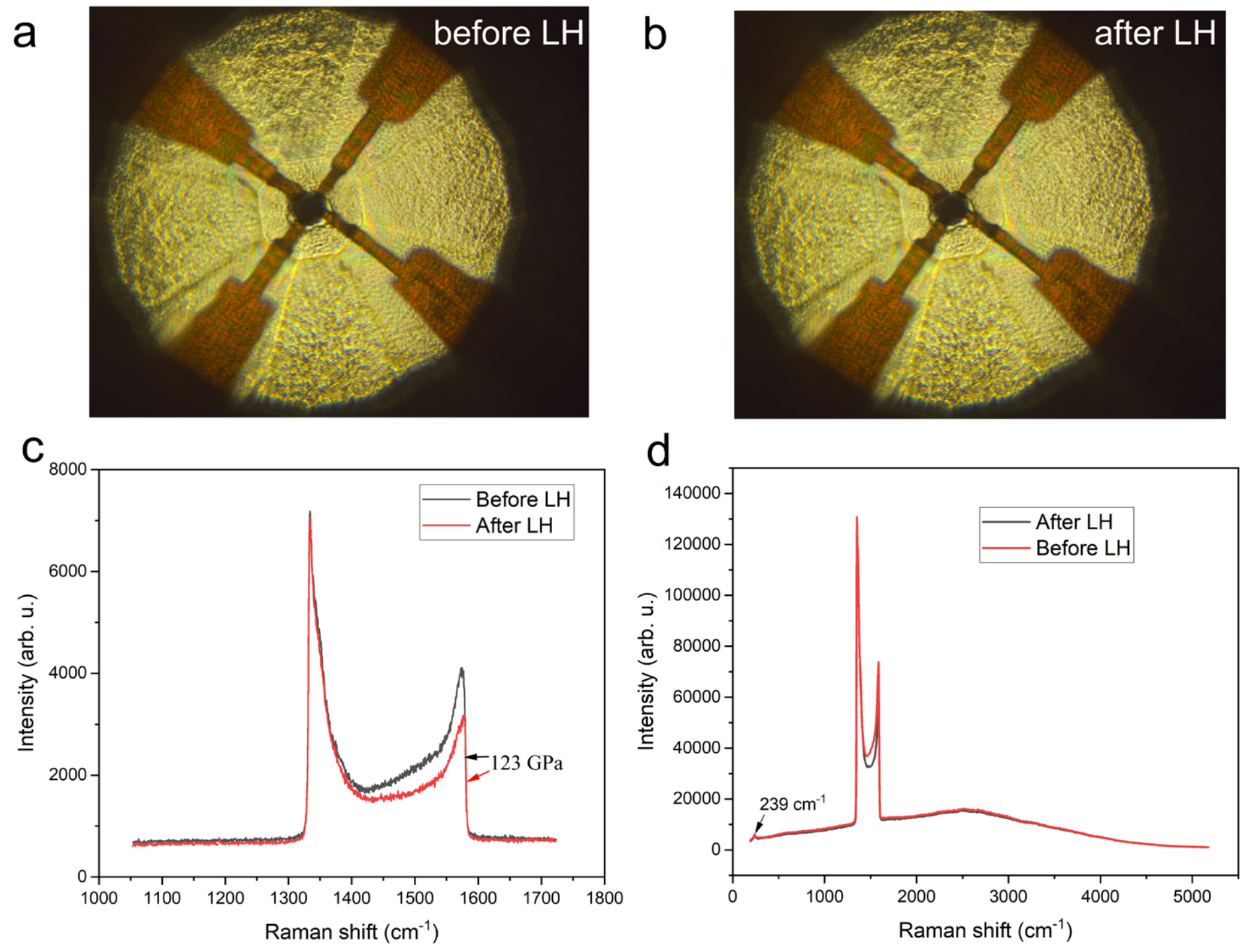

**Figure S6.** Photographs of La-Th hydride samples before and after additional (run 4) laser heating at 123 GPa, and Raman spectroscopy of the synthesized samples. (a, b) Photographs of the DAC's culet before and after additional (run 4) laser heating of the loaded $La_2Th$ particle. (c, d) Raman spectra of the center of high-pressure chamber before and after laser heating. A small signal at 239 cm$^{-1}$ is detected.

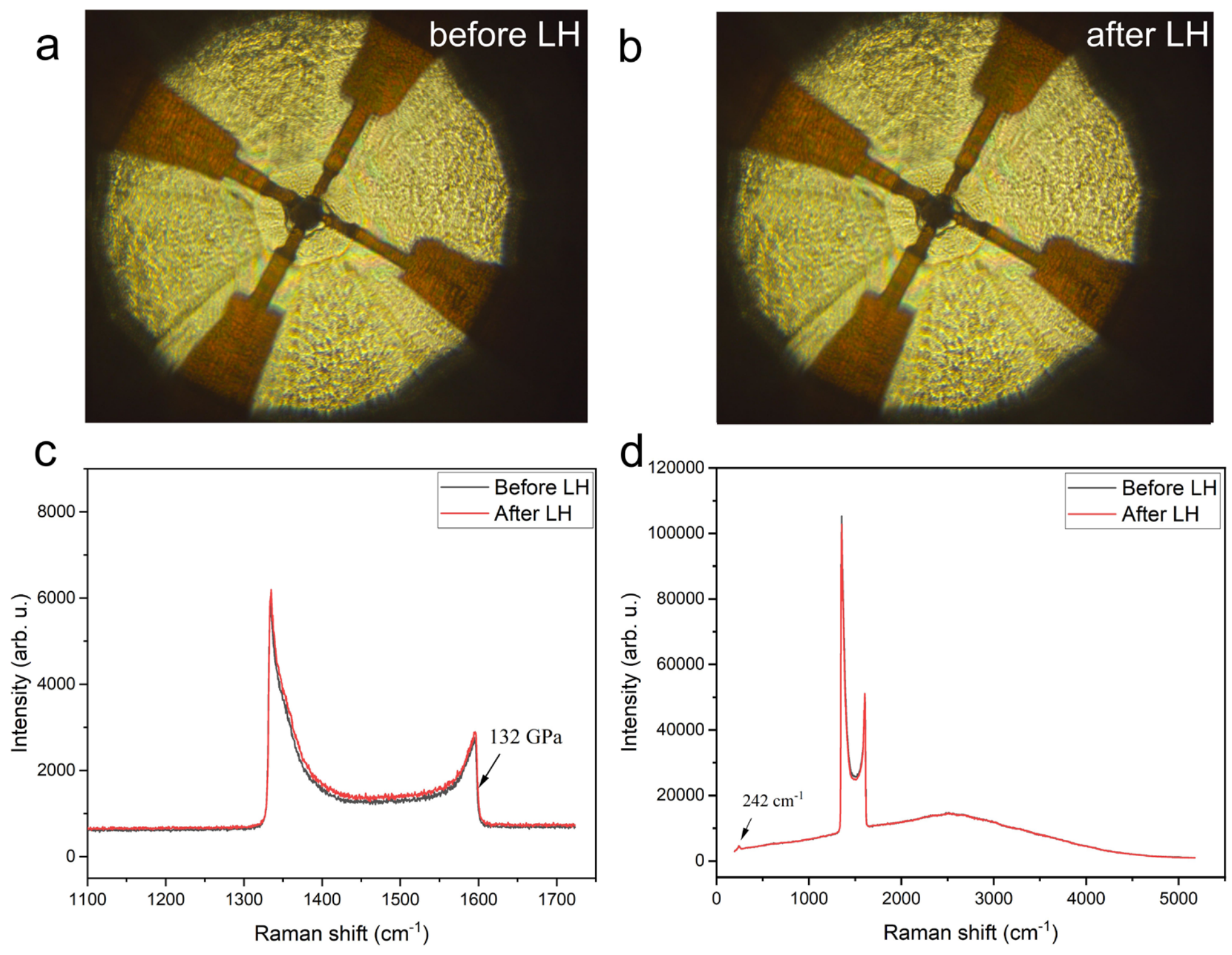

**Figure S7.** Photographs of La-Th hydride samples before and after additional (run 5) laser heating at 132 GPa, and Raman spectroscopy of the synthesized samples. (a, b) Photographs of the DAC's chamber before and after additional (run 5) laser heating of the loaded $La_2Th$ particle. (c, d) Raman spectra of the center of high-pressure cell before and after laser heating. A small signal at 242 $cm^{-1}$ is also detected.

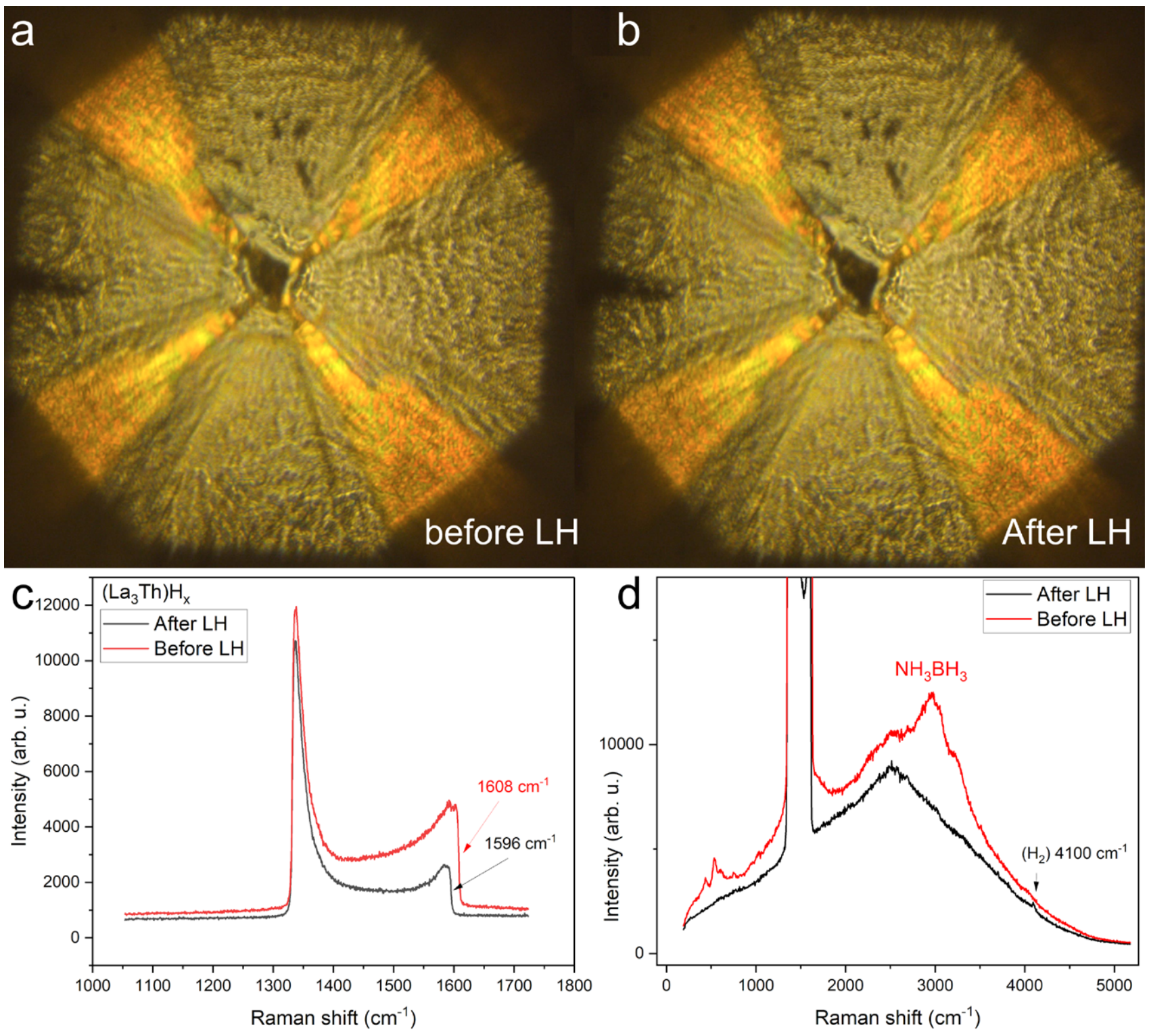

**Figure S8.** Photographs of $La_3Th$ hydride samples before and after laser heating at 137 GPa, and Raman spectra of a DAC's sample with the $La_3Th$/AB and corresponding hydrides before and after laser heating (LH). (a, b) Photographs of the DAC's chamber before and after laser heating. (c) Raman spectra in the range of 1000–1750 $cm^{-1}$, showing the diamond anvil peak edge at 1608 $cm^{-1}$ before laser heating (red) to 1596 $cm^{-1}$ after laser heating (black). (d) Full-range Raman spectra comparing the precursor contribution from ammonia borane ($NH_3BH_3$, red) before laser heating with the synthesized product after laser heating (black), highlighting the hydrogen ($H_2$) vibration peak near 4100 $cm^{-1}$.

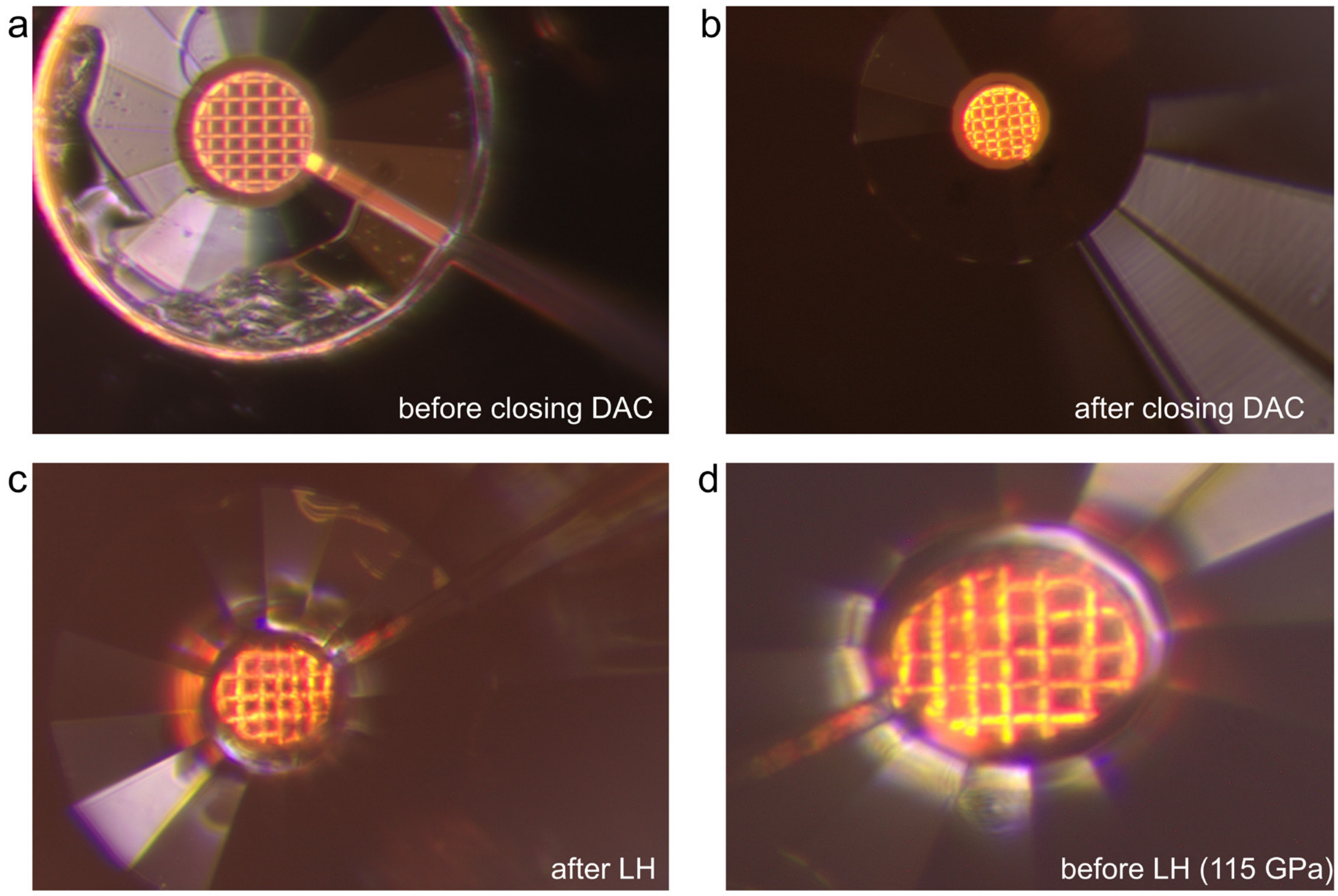

**Figure S9.** Optical photographs of preparation, closing, pressure buildup (~115 GPa) and laser heating of the DAC SM with matrix elements made of $La_2Th$ alloy. (a) Anvil with the sample before DAC's closure. (b) Sample after DAC's closure. (c) Sample after laser heating at 115-125 GPa. (d) Sample before laser heating.

## 4. X-ray diffraction measurements

In addition to the main phases $XH_{10}$ and $XH_9$, numerous other products of laser heating of La-Th alloys with $NH_3BH_3$ have been detected. One such byproduct is a hexagonal hydride with a small amount of hydrogen, *P*6/*mmm*-$XH_2$, whose diffraction peaks can be explained using two known prototypes: *P*6/*mmm*-$YH_2$ (Figures S15) and the recently proposed prototype *P*6/*mmm*-$Y_3H_{11}$ [19], which provides a much better explanation for the presence of small diffraction peaks (Figure 2a) and also better matches the calculated unit cell volume (V ~ 22 Å³/(La,Th) at 130 GPa).

A similar situation arises with the DAC SM: small-angle reflections at $2\theta$ = 6-7 deg. can be indexed using the *P*6/*mmm* group (as for *P*6/*mmm*-$LaH_{16}$ [20]), and in combination with *fcc*-(La,Th)$H_{10-x}$, all other reflections can be explained (Figures S11, S12). However, the volume of the cubic phase is less than expected for $XH_{10}$ at this pressure by 3-4 Å, which corresponds to the composition of $XH_8$. Another solution is to use *Fmmm*-(La,Th)$H_{18-x}$ in combination with (La,Th)$H_6$ (Figure S13). This allows for a better explanation of the peak intensities at $2\theta$ = 6-7 deg., as well as the presence of a radio-frequency anomaly near 253 K (Figure 4d).

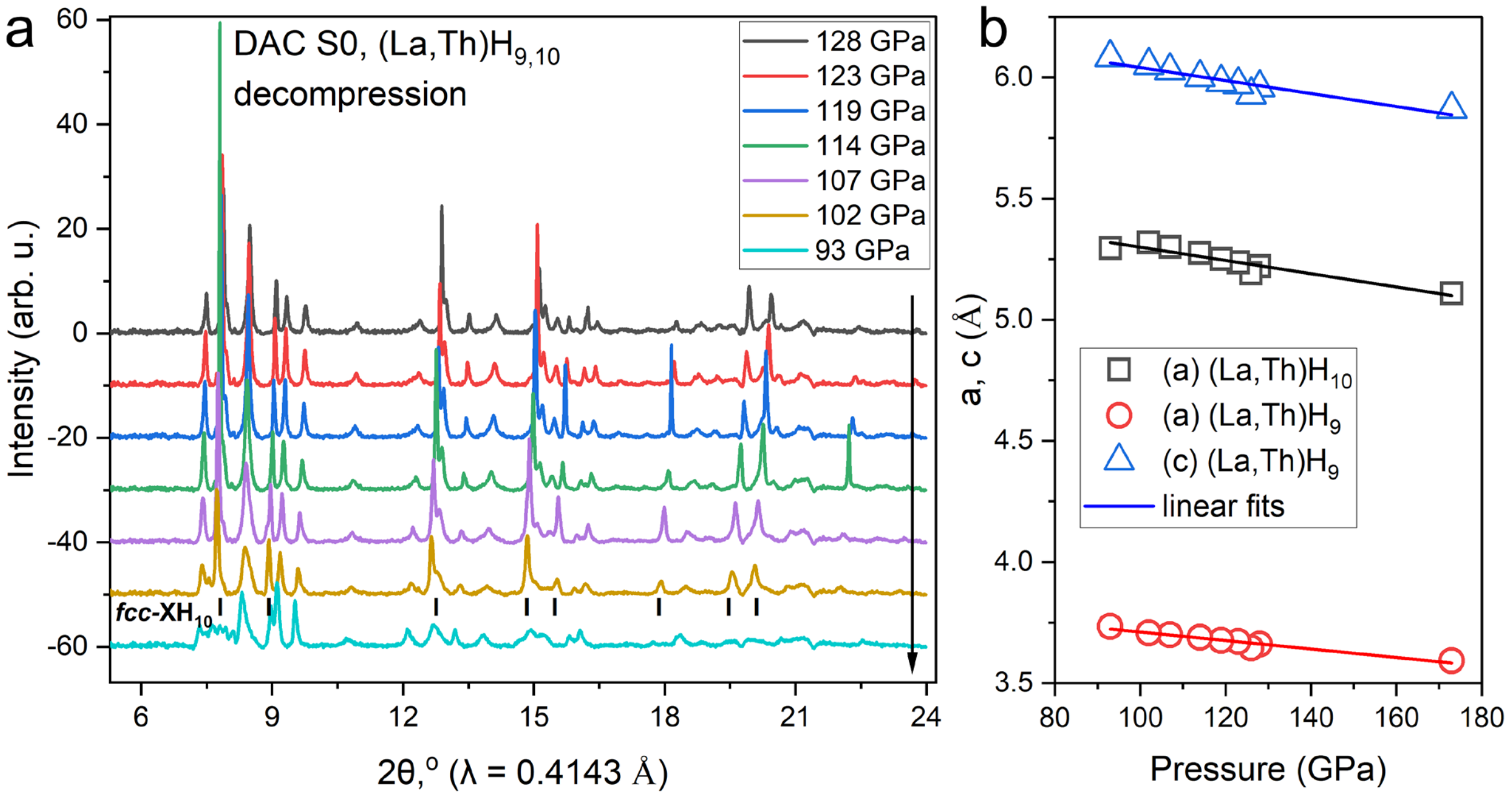

**Figure S10.** X-ray diffraction patterns and unit cell parameters of (La,Th)$H_{9-10}$ during decompression in DAC S0. (a) Evolution of XRD patterns with decreasing pressure from 128 GPa to 93 GPa. The evolution of XRD peaks corresponding to the face-centered cubic (*fcc*) (La,Th)$H_{10}$ phase are highlighted. (b) Pressure dependence of the unit cell parameters for cubic (La,Th)$H_{10}$ (black squares), and for hexagonal (La,Th)$H_9$ (red circles and blue triangles, respectively). Linear fits to the data illustrate the trend of the lattice parameters changing with pressure. Data points at 173 GPa belong to DAC S3.

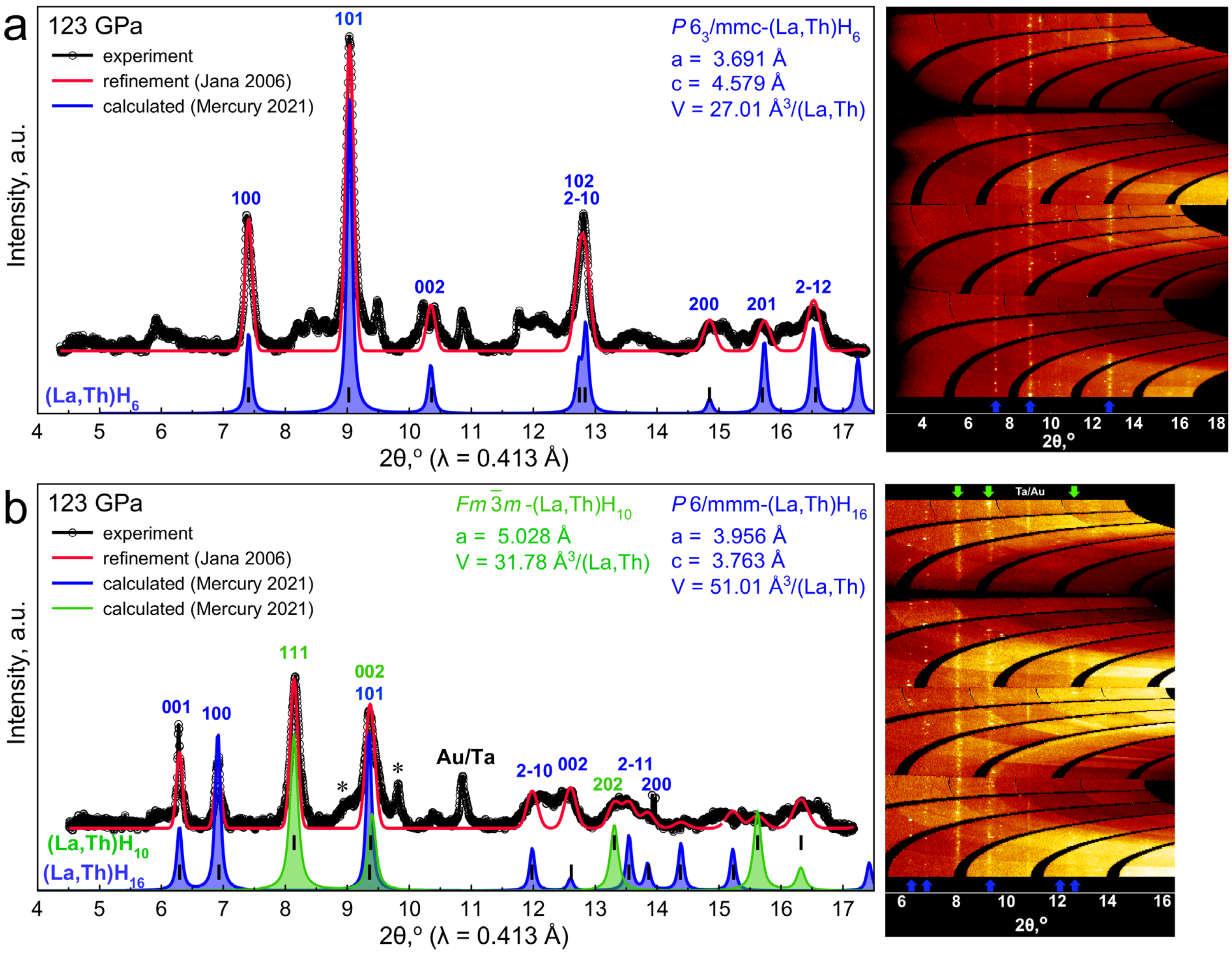

**Figure S11.** Synthesis of La-Th polyhydrides in DAC SM at 123 GPa. (a) Experimental XRD pattern and Le Bail refinement of the hexagonal *P*6$_3$/*mc*- or *P*6$_3$/*mmc*-(La,Th)$H_6$ phase unit cell parameters. The experimental data (open circles), refinement (solid red line), and calculated XRD (solid blue line) are shown. The refined lattice parameters and unit cell volume are provided in the legend. The corresponding raw XRD image ("cake") is shown on the right panel (wavelength is 0.413 Å). (b) Experimental XRD pattern and Le Bail refinement of a mixture of cubic *Fm*$\bar{3}$*m*-(La,Th)$H_{10-x}$ and hexagonal *P*6/*mmm*-(La,Th)$H_{16}$ phases. The green and blue solid lines represent the calculated XRD patterns for the respective phases. The asterisks mark unexplained peaks, and the reflections from Au/Ta lenz are labeled. The corresponding raw XRD image ("cake") is shown on the right panel.

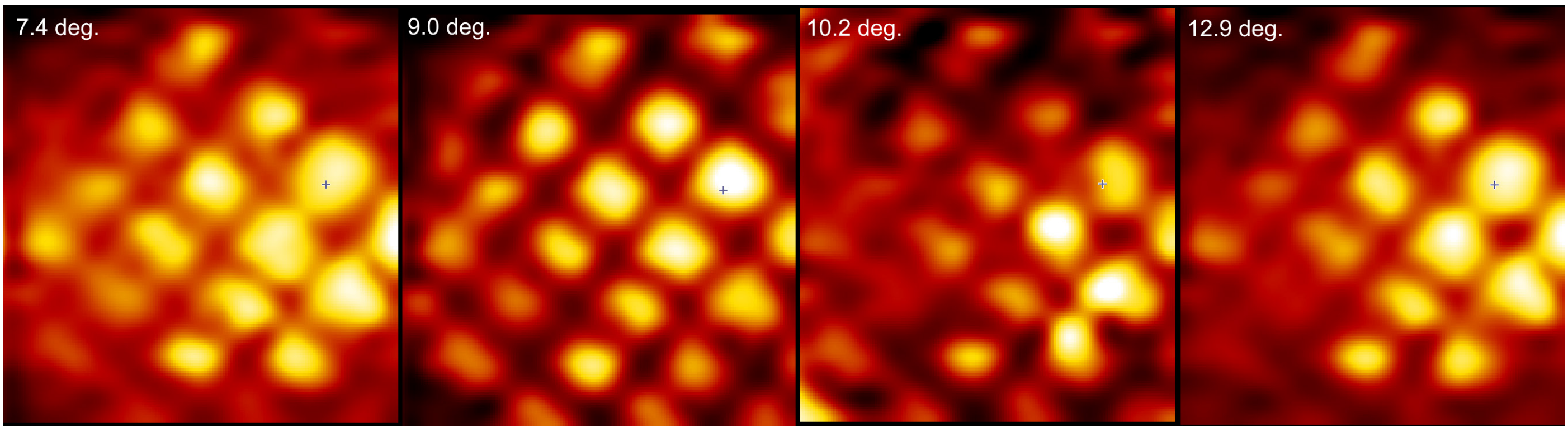

**Figure S12.** Heat colormaps of the spatial distribution of the XRD intensity of the four diffraction reflections of the hexagonal phase *P*6$_3$/*mmc*-(La,Th)$H_6$ in DAC SM at 123 GPa, showing their correlation. Image scale: 60 μm.

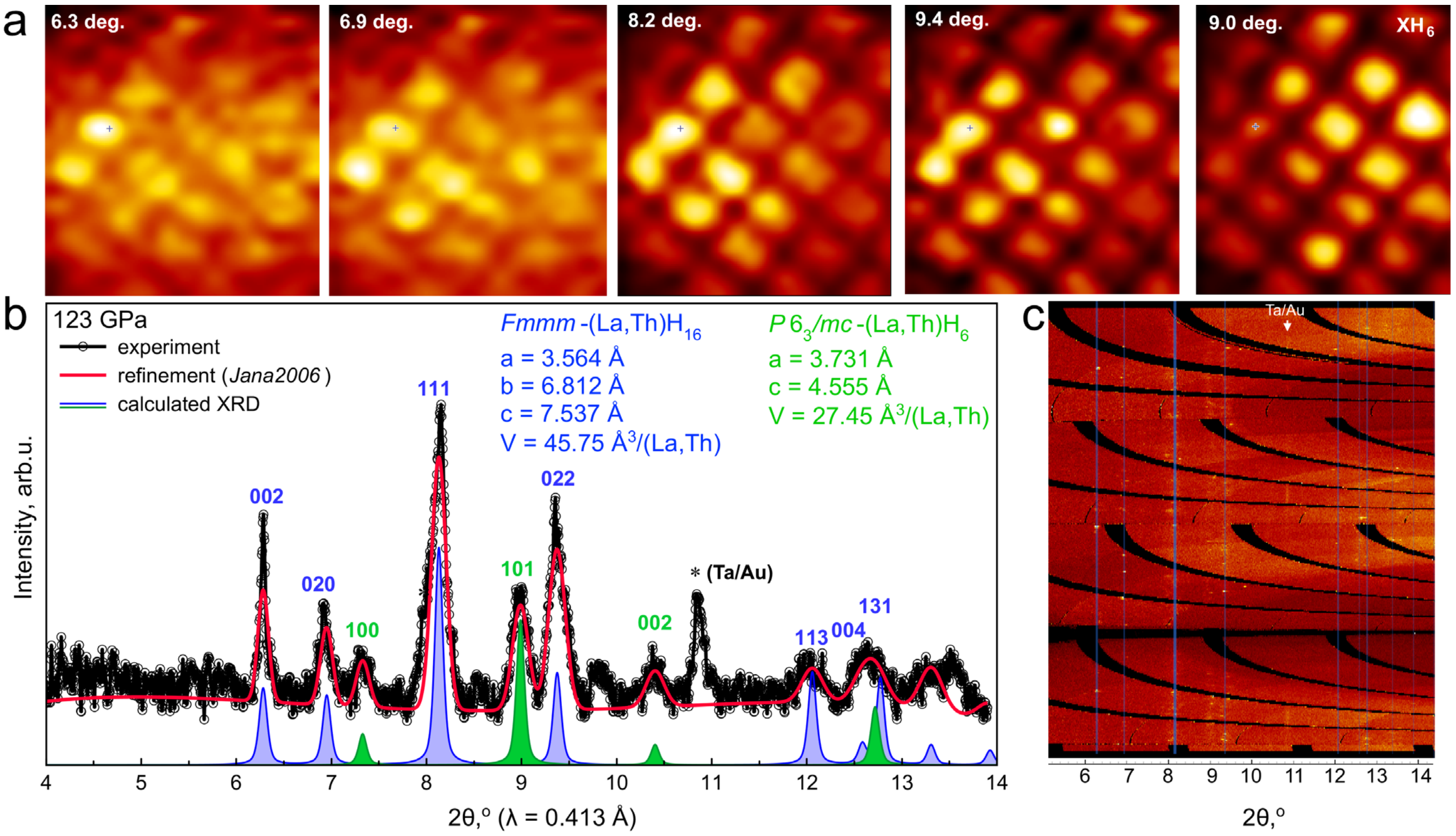

**Figure S13.** Experimental diffraction pattern of DAC SM's sample at 123 GPa and Le Bail refinement of the lattice parameters of the *Fmmm*-$XH_{16}$ and *P*$6_3$/*mc*-$XH_6$ models used to explain the experimental reflections. (a) Series of heat maps of the spatial distribution of XRD intensity for the four first diffraction reflections of the *Fmmm*-$XH_{16}$ phase showing their correlation in the left corner of the sample compared with the distribution of the hexagonal $XH_6$ phase (9.0 deg.), concentrated in the right corner of the sample. (b) Le Bail refinement. The experimental data (open circles), refinement (solid red line), and calculated XRD patterns (solid blue and green curves) are shown. Unexplained narrow XRD peak corresponds to the Ta/Au Lenz lens. (c) The raw XRD image ("cake") with blue vertical lines corresponding to the *Fmmm* phase is shown on the right panel (wavelength is 0.413 Å).

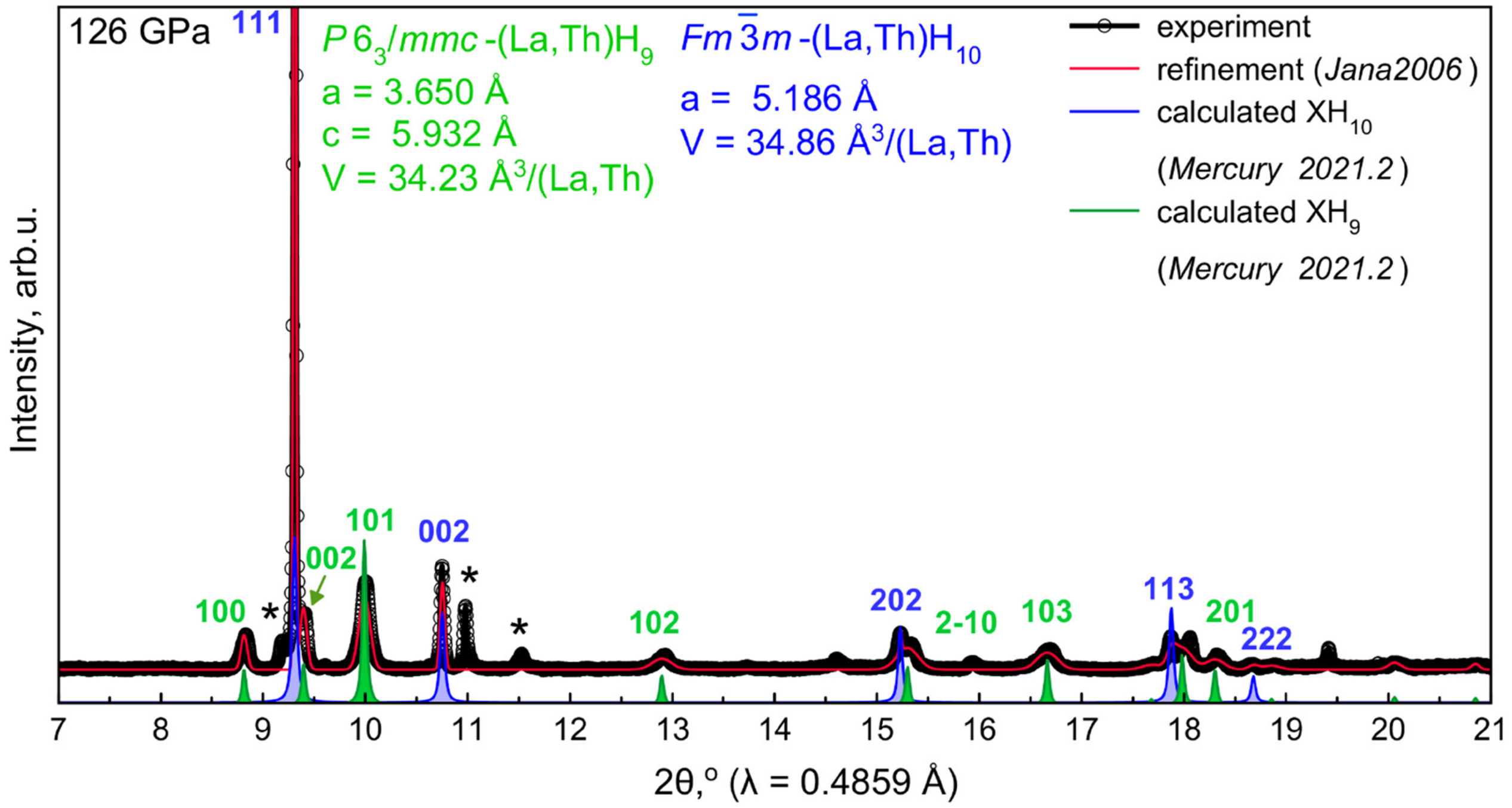

**Figure S14.** Experimental XRD and the Le Bail refinement of cubic and hexagonal $(La,Th)H_{9-10}$ unit cells at 126 GPa (DAC S1) using an X-ray wavelength of 0.4859 Å. The experimental profile is shown as a black line with circles, and the refinement (Jana2006) is indicated by the red line. Calculated XRD peaks for the cubic *Fm*$\bar{3}$*m*-$(La,Th)H_{10}$ phase are shown in blue (lattice parameter a = 5.186 Å, unit cell volume per metal atom V = 34.86 Å$^3$/(La,Th)). Calculated XRD peaks for the *P*$6_3$/*mmc*-$(La,Th)H_9$ phase are shown in green (lattice parameters a = 3.650 Å, c = 5.932 Å, volume V = 34.23 Å$^3$/(La,Th)). Miller indices (hkl) for each phase are color-coded accordingly, and unidentified impurity (likely *P*6/*mmm*-$(La,Th)H_{2-x}$) reflections are marked with asterisks (*).

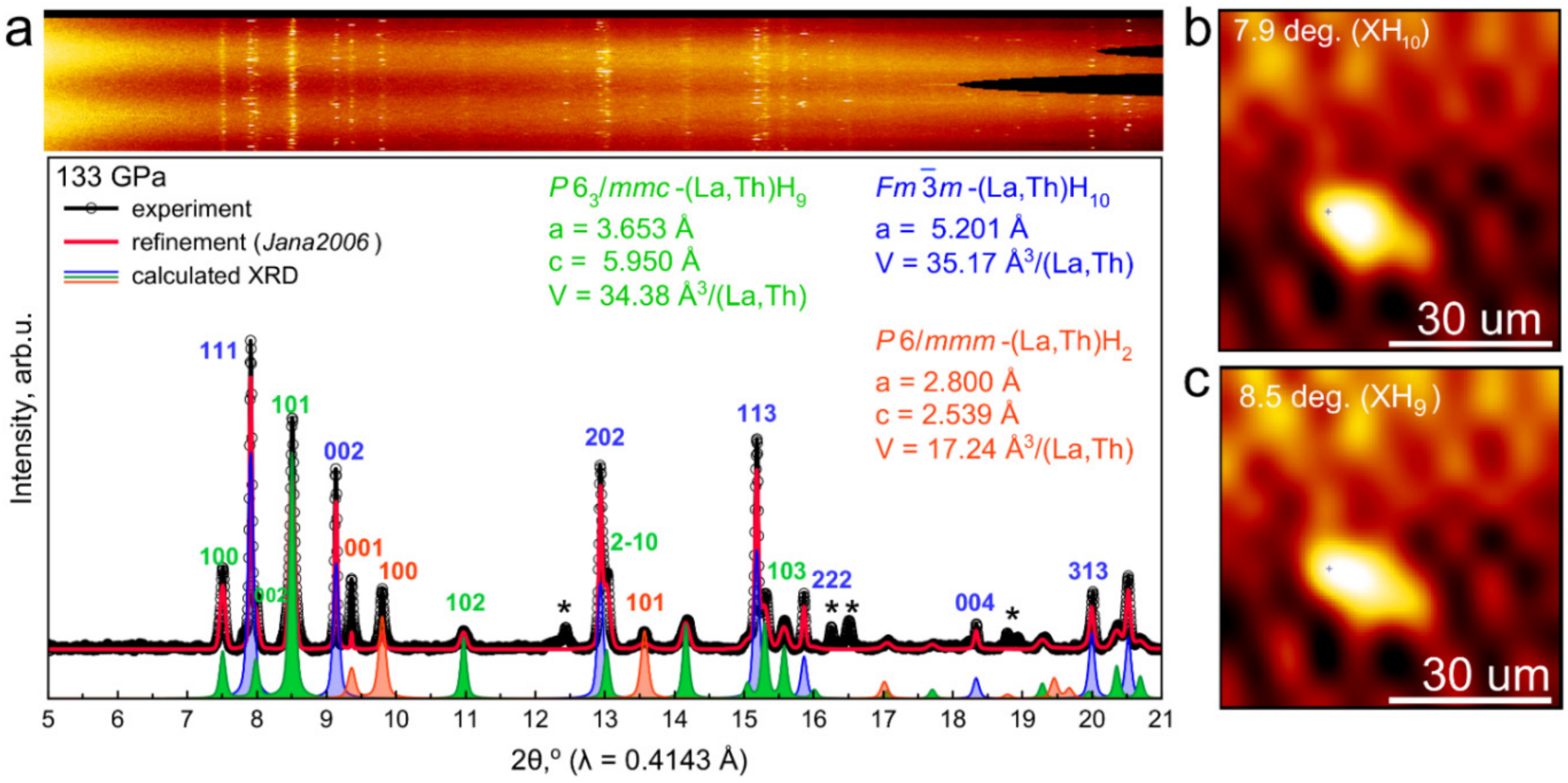

**Figure S15.** Experimental XRD and the Le Bail refinement of cubic and hexagonal unit cells of La-Th hydrides at 133 GPa in DAC S1. (a) The 2D diffraction image ("cake", top) and the integrated 1D experimental profile (bottom) shown as a black line with circles, with the refinement (Jana2006) indicated by the red line. Calculated XRD peaks for the cubic (La,Th)$H_{10}$ phase are shown in blue (lattice parameter a = 5.201 Å, unit cell volume per metal atom V = 35.17 Å$^3$/(La,Th)). Calculated XRD pattern for the hexagonal *P*6$_3$/*mmc*-(La,Th)$H_9$ phase is shown in green (lattice parameters a = 3.653 Å, c = 5.950 Å, volume V = 34.38 Å$^3$/(La,Th)). Calculated XRD pattern for possible *P*6/*mmm*-(La,Th)$H_{2-x}$ phase is shown in orange. Miller indices (hkl) for each phase are color-coded accordingly, and impurity reflections are marked with asterisks (*). (b, c) X-ray mapping images (intensity distribution) at 2θ = 7.9° ($XH_{10}$) and 8.5° ($XH_9$), respectively, with a scale bar of 30 um.

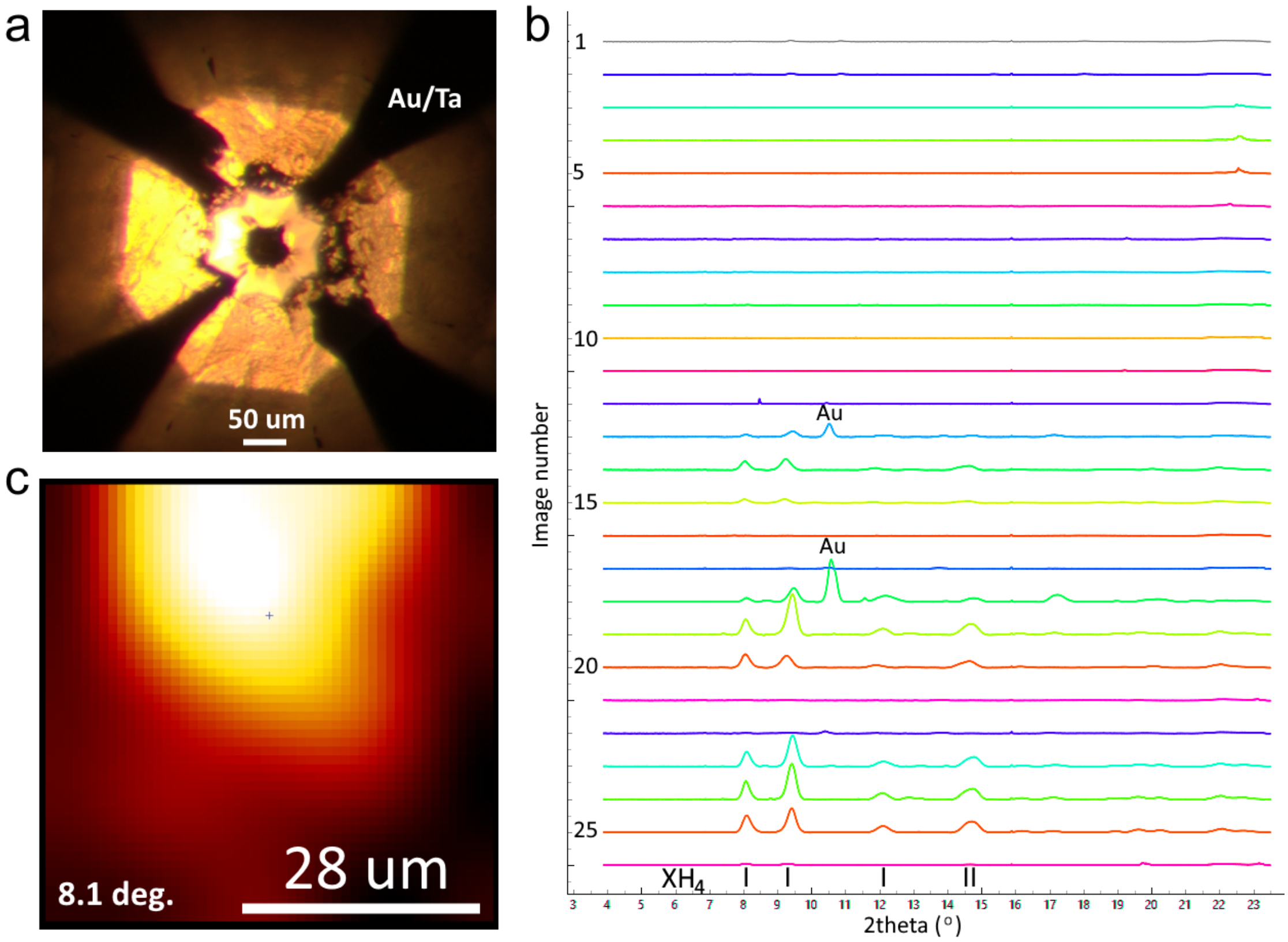

**Figure S16.** Synthesis of *I*4/*mmm*-(La,Th)$H_4$ in DAC S4. (a) Optical photograph of the DAC S4's culet after laser heating with "spitting" of the $CaF_2$/epoxy insulating gasket and damage to the electrodes at 174-180 GPa. As a result of $CO_2$ laser heating, there was a pressure jump from 174 to 180 GPa, and a uniform sample of fine-grained powdered (La,Th)$H_4$ was formed, the diffraction patterns of which are shown in panel (b). (b) XRD patterns corresponding to image numbers 1 to 26 (5×5 mapping, 11 µm step, beam diameter was about 5 µm), showing the formation of the target (La,Th)$H_4$ tetrahydride along with residual material from the Au electrodes. (c) Thermal map of the distribution of the XRD peak at 8.1 deg. corresponding to the target tetrahydride.

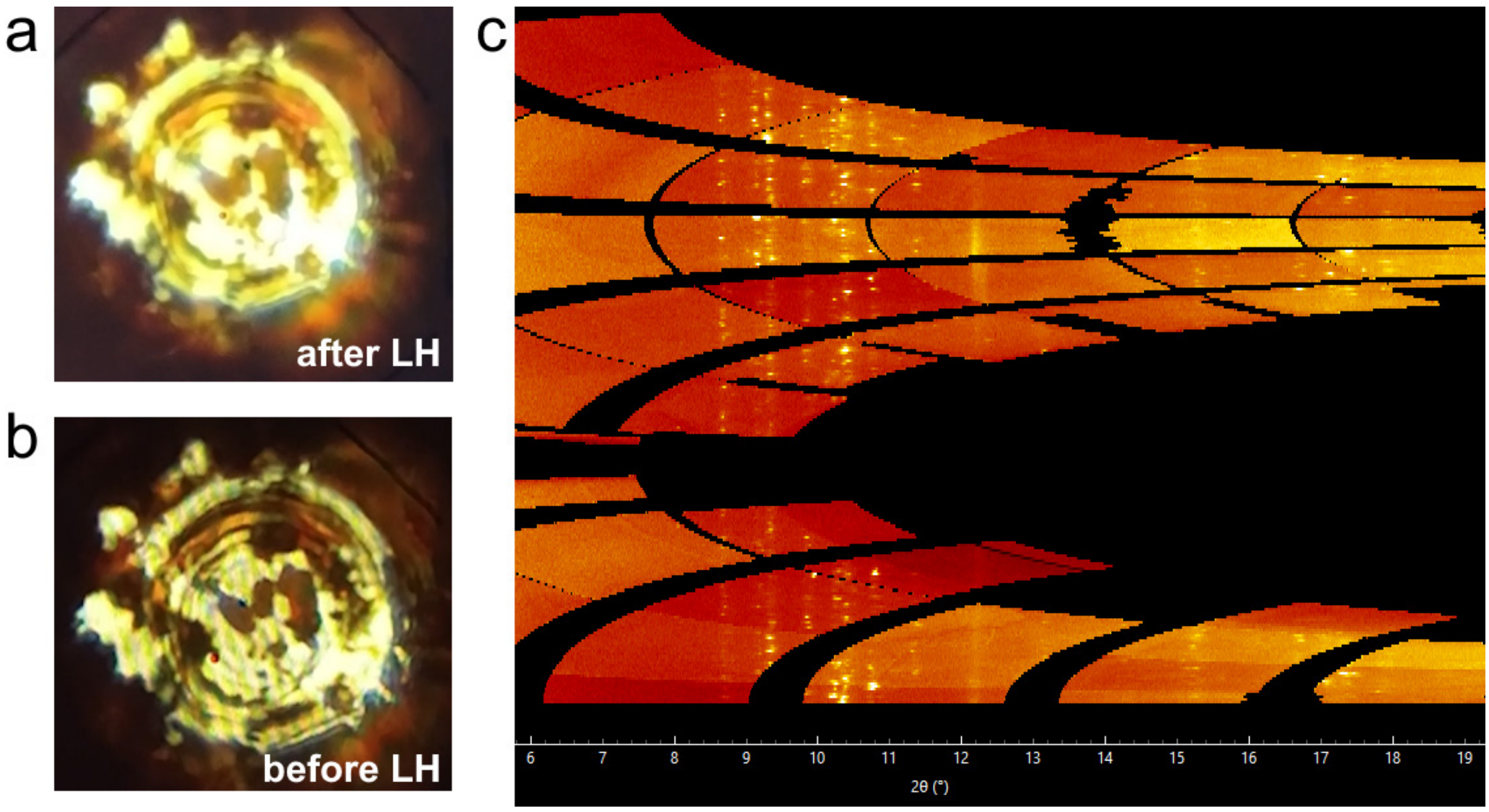

**Figure S17.** Synthesis of La-Th polyhydrides at 70 GPa from $La_5Th$ alloy. (a, b) Optical photographs of the DAC S5's culet with the sample at 70 GPa (a) after and (b) before the laser heating (1 μm laser, 0.3 s pulses). After laser heating, two of the three $La_5Th$ particles effectively fused into one. (c) Raw XRD image ("cake") showing the formation of the $X_{13}H_{75}$ phase at 70 GPa (wavelength is 0.496 Å).

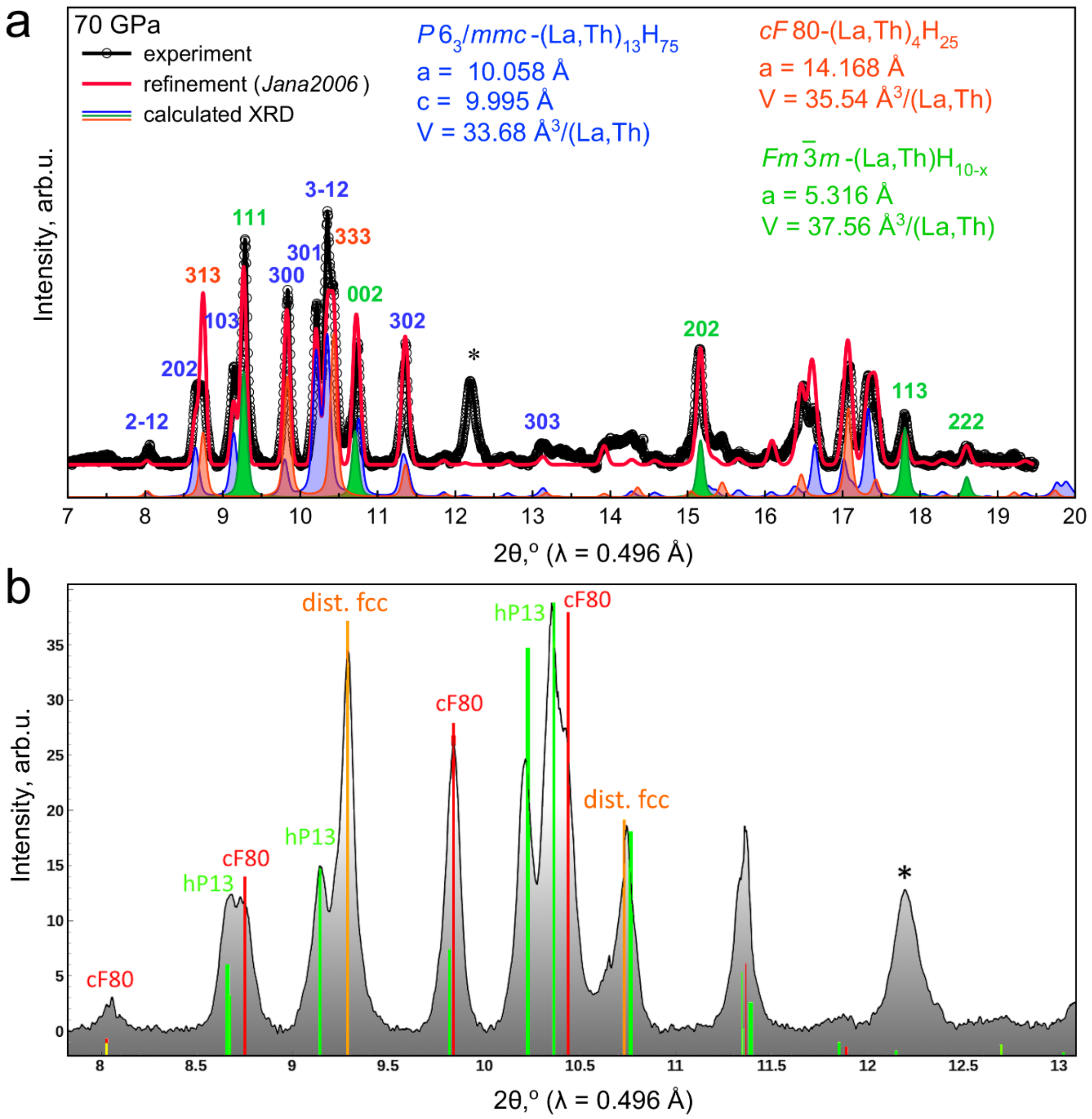

**Figure S18**. Synthesis of La-Th polyhydrides at 70 GPa in DAC S5. (a) Experimental XRD and Le Bail refinement of the unit cell parameters of *cF*80, *hP*13, and dist. *fcc* phases. The experimental data (open circles), refinement profile (solid red line, profile was not refined), and calculated XRD patterns for the identified phases are shown. The blue, red, and green curves indicate the calculated XRD for the $P6_3/mmc$-(La,Th)$_{13}$H$_{75}$, *cF*80-(La,Th)$_4$H$_{25}$, and $Fm\bar{3}m$-(La,Th)H$_{10-x}$ phases, respectively. The corresponding refined lattice parameters and unit cell volumes are provided in the legend. (b) Qualitative comparison of the experimental XRD profile and predicted diffraction peaks for the *cF*80 (red), *hP*13 (green), and dist. *fcc* (orange) phases, which are marked by vertical lines. The asterisk in both panels marks an unexplained diffraction peak. The wavelength is 0.496 Å.

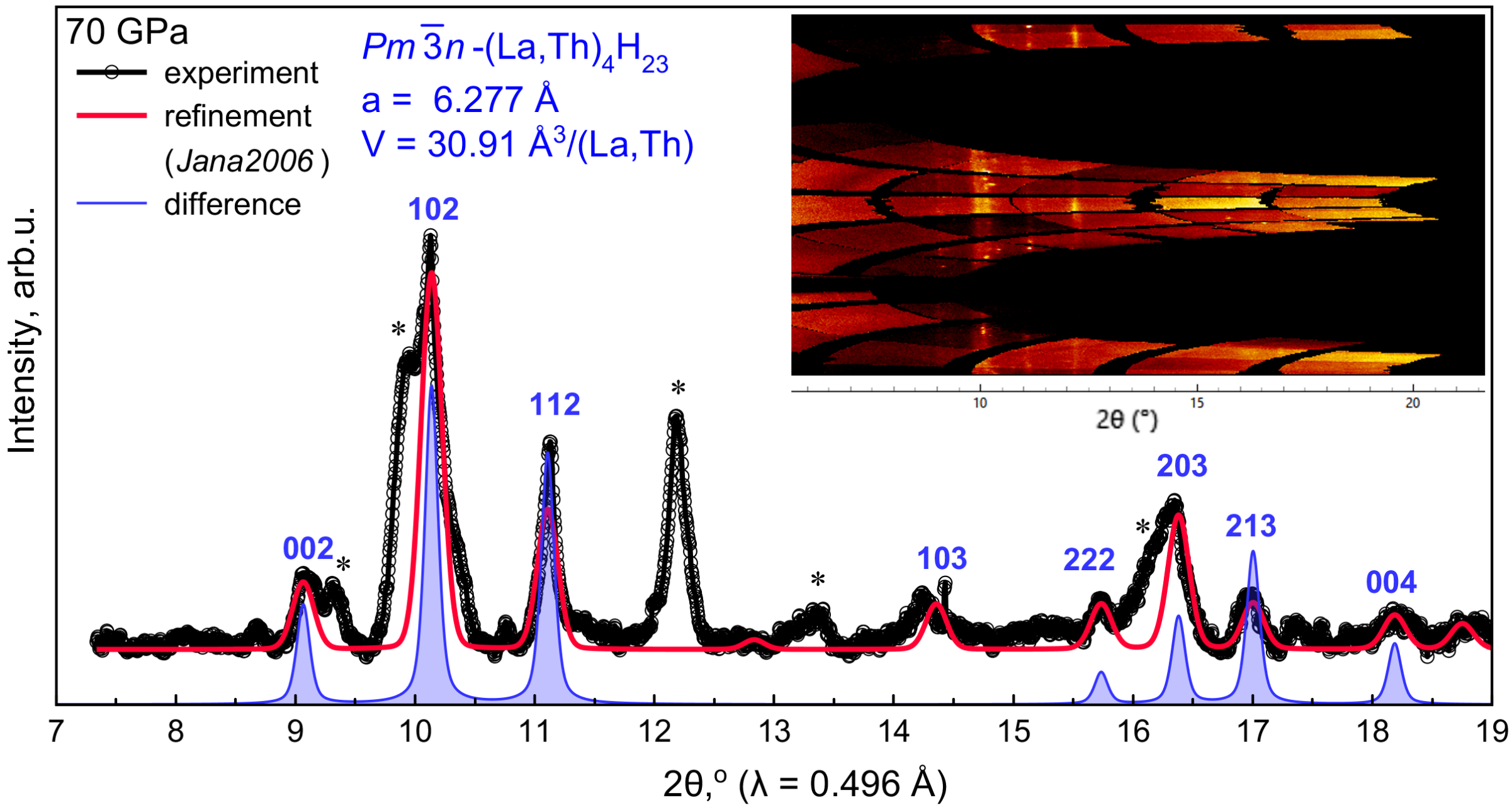

**Figure S19.** Synthesis of A15-$(La,Th)_4H_{23}$ polyhydride at 70 GPa. (a) Experimental XRD and Le Bail refinement of the A15 phase unit cell parameter. The experimental data (black open circles), refinement profile (solid red line), and calculated XRD pattern (solid blue line) for the $Pm\bar{3}n$-$(La,Th)_4H_{23}$ are shown. The asterisks mark peaks of impurity phases. (b) Raw XRD image ("cake") showing the experimental diffraction pattern (wavelength is 0.496 Å).

**Table S2.** Unit cell parameters (in Å) and volumes (in Å$^3$/(La,Th)) for different La-Th polyhydrides identified by powder XRD diffraction in DACs tested at one pressure point. The accuracy of all parameters is two significant digits after the decimal point. Similar cell parameters obtained from different points of the same DAC and at the same pressure are not given here.

| ***P*(GPa)** | ***Fmmm*-$XH_{16}$** | ***fcc*-$XH_{10}$** | ***hcp*-$XH_9$** | ***hP* or *Cmc*2$_1$-$XH_6$** | ***cF*80-$X_4H_{25}$** | ***hP*13-$X_{13}H_{75}$** | **A15 $X_4H_{23}$** | ***I4/mmm*-$XH_4$** | ***hP*3-$X_3H_{11}$** |
|---|---|---|---|---|---|---|---|---|---|
| 70 | | a = 5.32<br>V=37.56 | | | a=14.17<br>V=35.5 | a = 10.06<br>c = 9.99<br>V= 33.68 | a = 6.28<br>V=30.91 | | |
| 123 | a = 3.56<br>b = 6.81<br>c = 7.54<br>V=45.75 | a = 5.03<br>V=31.78 | | a = 3.73<br>c = 4.55<br>V=27.45 | | | | | |
| 126 | | a = 5.19<br>V=34.86 | a = 3.65<br>c = 5.93<br>V=34.23 | | | | | | a=5.08<br>c=2.90<br>V=21.55 |
| 133 | | a=5.20<br>V=35.17 | | | | | | | |
| 173 | | a=5.11<br>V=33.3 | a=3.59<br>c=5.87<br>V=32.8 | a=3.47<br>b=5.64<br>c=5.48<br>V=26.9 | | | | | |
| 180 | | | | | | | | a = 2.78<br>c = 5.87<br>V= 22.76 | |

## 5. Additional transport data

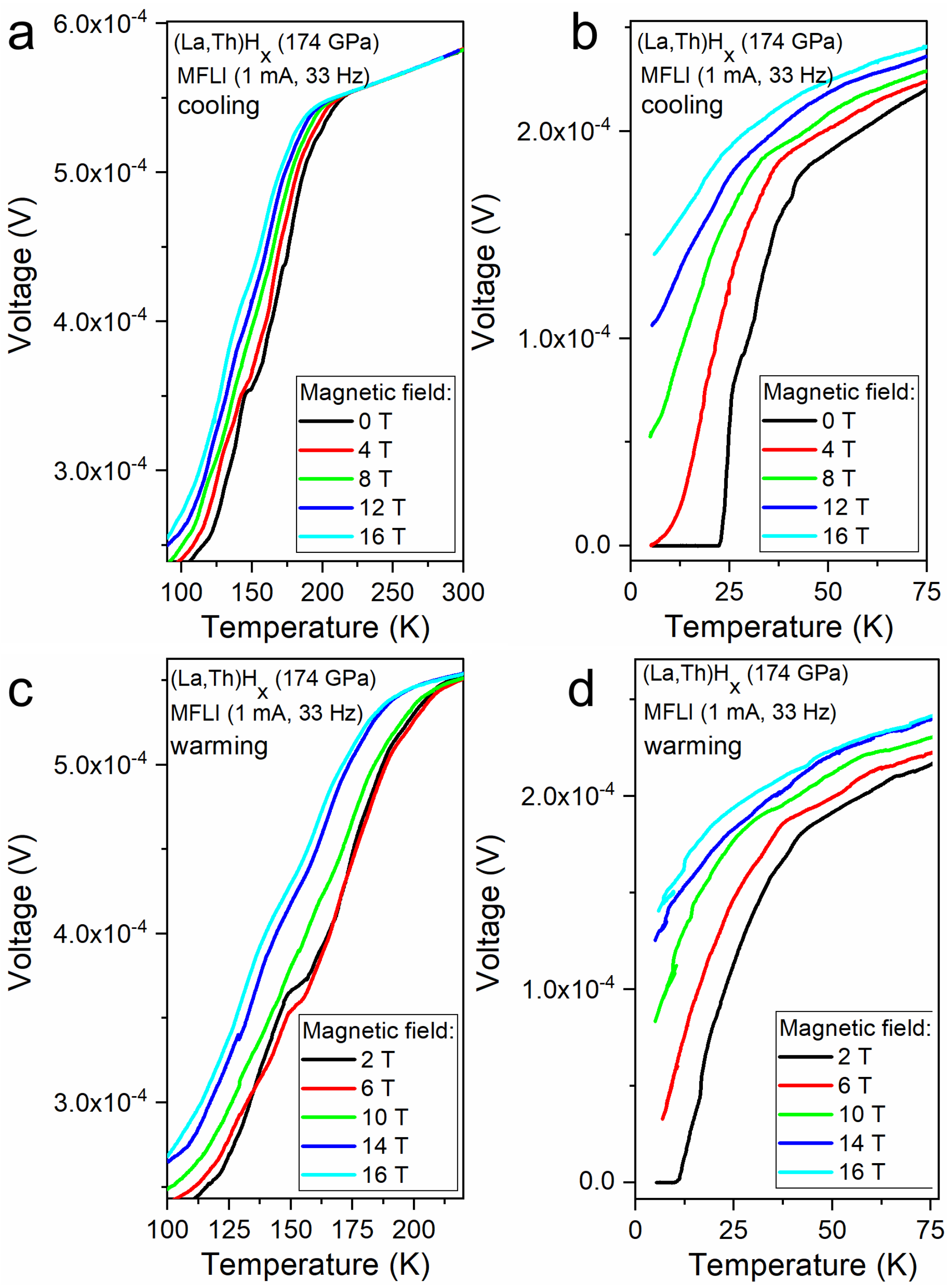

**Figure S20.** Temperature dependence of real part of AC voltage drop across the (La,Th)$H_{9-10}$ sample measured at a pressure of 173 GPa (DAC S3) under various external magnetic fields, recorded using a lock-in amplifier (MFLI, current 1 mA, frequency 33 Hz). (a) Main temperature range (100-300 K) during cooling under applied magnetic fields of 0 T, 4 T, 8 T, 12 T, and 16 T. (b) Magnified low-temperature region (0-75 K) with low-$T_c$ superconducting phases ($XH_4$ ? See Ref. [21]) studied upon cooling, highlighting the suppression of the superconducting transition in with increasing magnetic fields from 0 T to 16 T. (c) Temperature-dependent voltage during warming over the range of 100-225 K measured at magnetic fields of 2 T, 6 T, 10 T, 14 T, and 16 T. (d) Magnified low-temperature region (0-75 K) upon warming, demonstrating the magnetic field suppression of the transition state during the warming cycle.

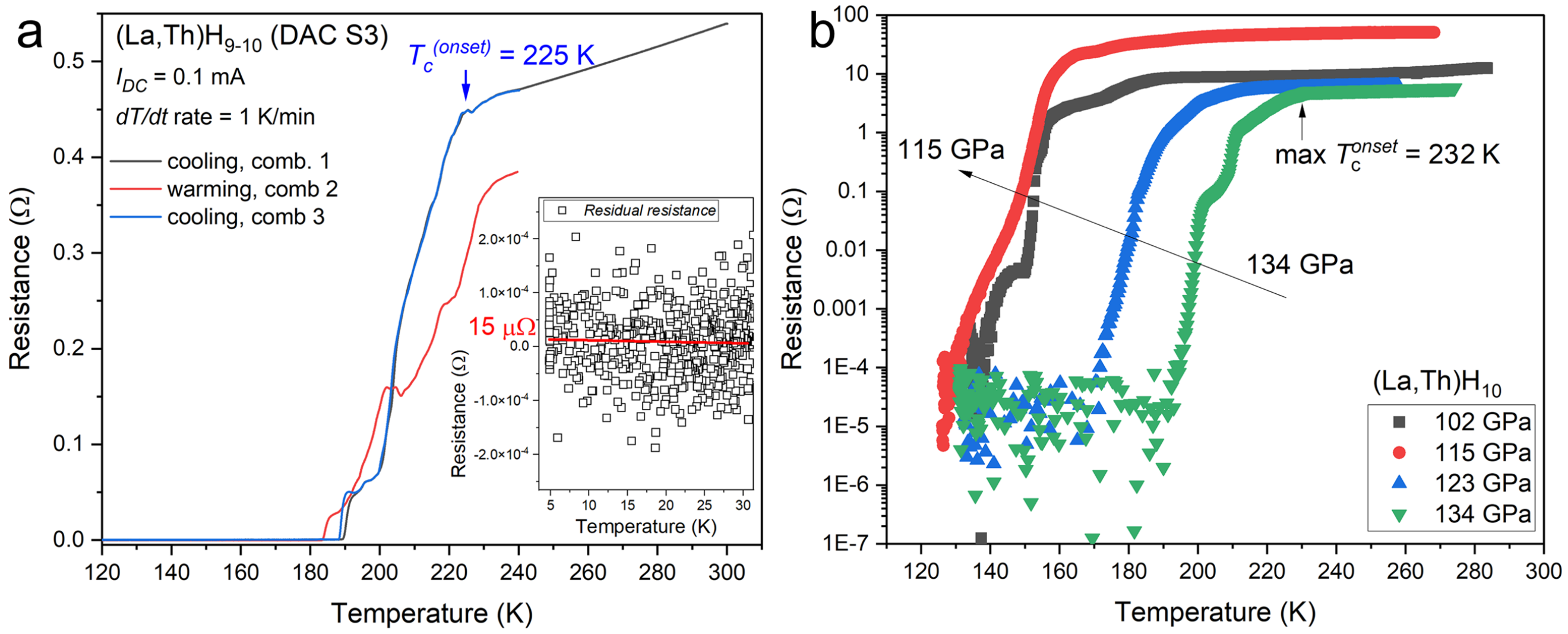

**Figure S21**. Transport properties of (La,Th)$H_{9\text{-}10}$ in the DACs S1 and S3. (a) Temperature dependence of electrical resistance of DAC S3's sample for three van der Pauw channels. We used delta mode and DC current of 0.1 mA. The inset shows the residual resistance of the sample, approximately 15 μΩ. (b) Logarithmic plot of resistance as a function of temperature for (La,Th)$H_{9\text{-}10}$ at various pressures (102 GPa, 115 GPa, 123 GPa, and 134 GPa), illustrating the sharp drop in resistance associated with the superconducting transition.

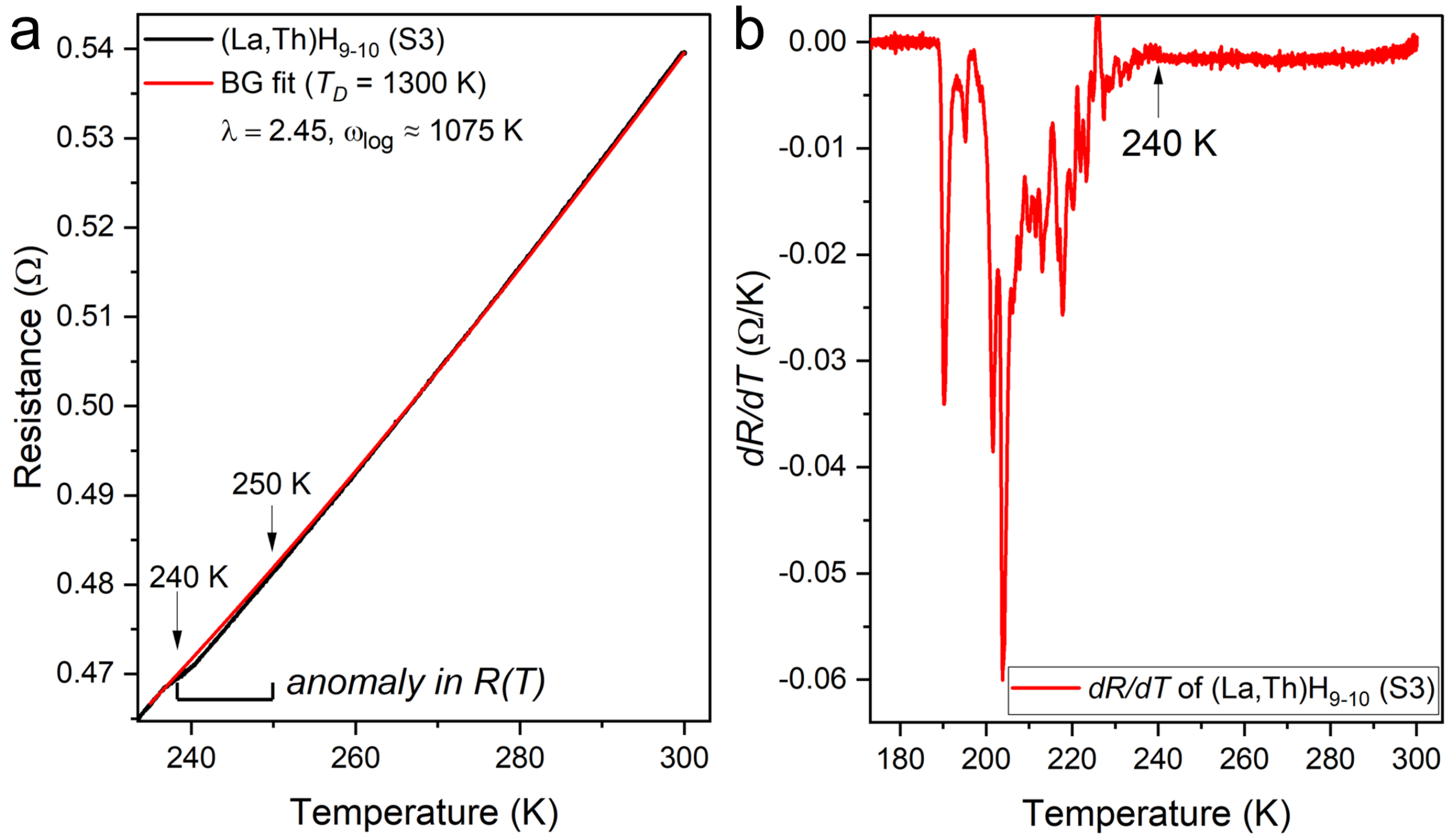

**Figure S22**. Study of electrical resistance in a DAC S3 sample at 173 GPa. (a) Anomaly in the electrical resistance of DAC S3 sample below 250 K, and a Bloch-Grüneisen fit of the *R* vs *T* dependence. This anomaly may correspond to the appearance of superconductivity in some hydride (La,Th)$H_x$. (b) First temperature derivative of the electrical resistance *dR/dT* for this sample, indicating the onset of superconductivity starting from 240 K at 173 GPa.

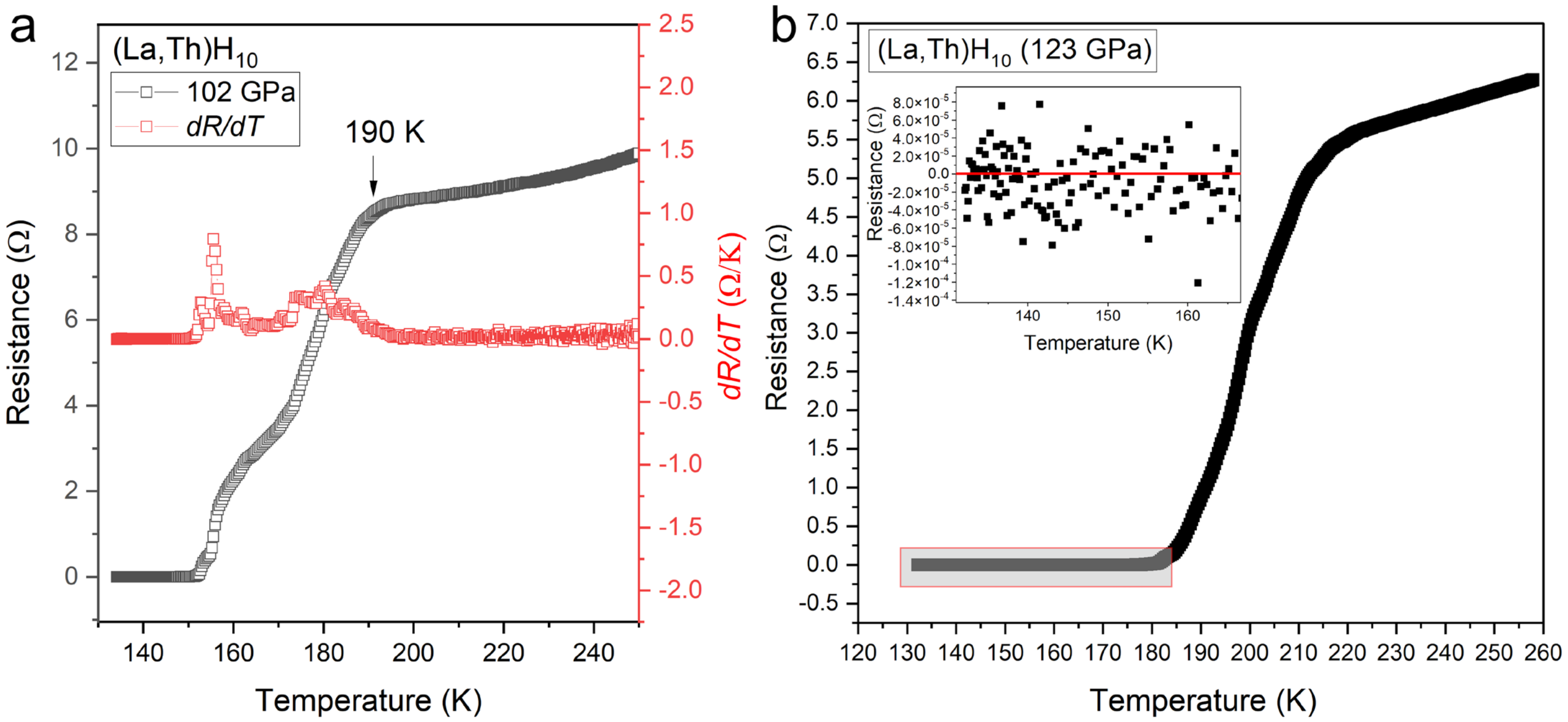

**Figure S23.** Temperature dependence of electrical resistance of $(La,Th)H_{9\text{-}10}$ hydrides synthesized under pressure of (a) 102 GPa and (b) 123 GPa. The panel (a) also shows the first derivative with respect to temperature *dR/dT*, which allows estimating the temperature of the onset of the superconducting transition ($T_c^{onset}$) as 195-200 K. The inset of panel (b) shows the residual resistance in the superconducting state.

**Supporting Table S3.** Parameters of the superconducting state of $(La,Th)H_9$ and $(La,Th)H_{10}$ at 173 GPa.

| **Parameter** | **$(La,Th)H_{10}$** | **$(La,Th)H_9$** |
|---|---|---|
| λ | 2.45 [a] | 2.55 [b] |
| $\omega_{log}$, K | 1075 [a] | 707 [b] |
| $T_c$, K | 225 (onset) | 190 (onset)<br>175 (offset) |
| Δ(0), meV | ≈ 51 [c] | 37 |
| $B_{c2}(0)$, T | 100–120 | 44 |
| $R_\Delta = 2\Delta(0)/k_BT_c$ | ≈5.26 [c] | 4.9 |
| $V_F$, m/s | $4.1–4.9\times10^5$ | $4.8\times10^5$ |
| $\lambda_L(0)$, nm | 17–31 [d] | 220 |
| ξ(0), nm | 1.65 | 2.73 |
| Ginzburg–Landau parameter (κ) | 10–20 [d] | 80 |
| Critical current density ($J_c$), A/m$^2$ | – | $1.95\times10^{10}$ |

[a] At pressure of 173 GPa; [b] DFT estimations at 180 GPa; [c] Formula eq. (1);
[d] Estimations based on Hall effect measurements.

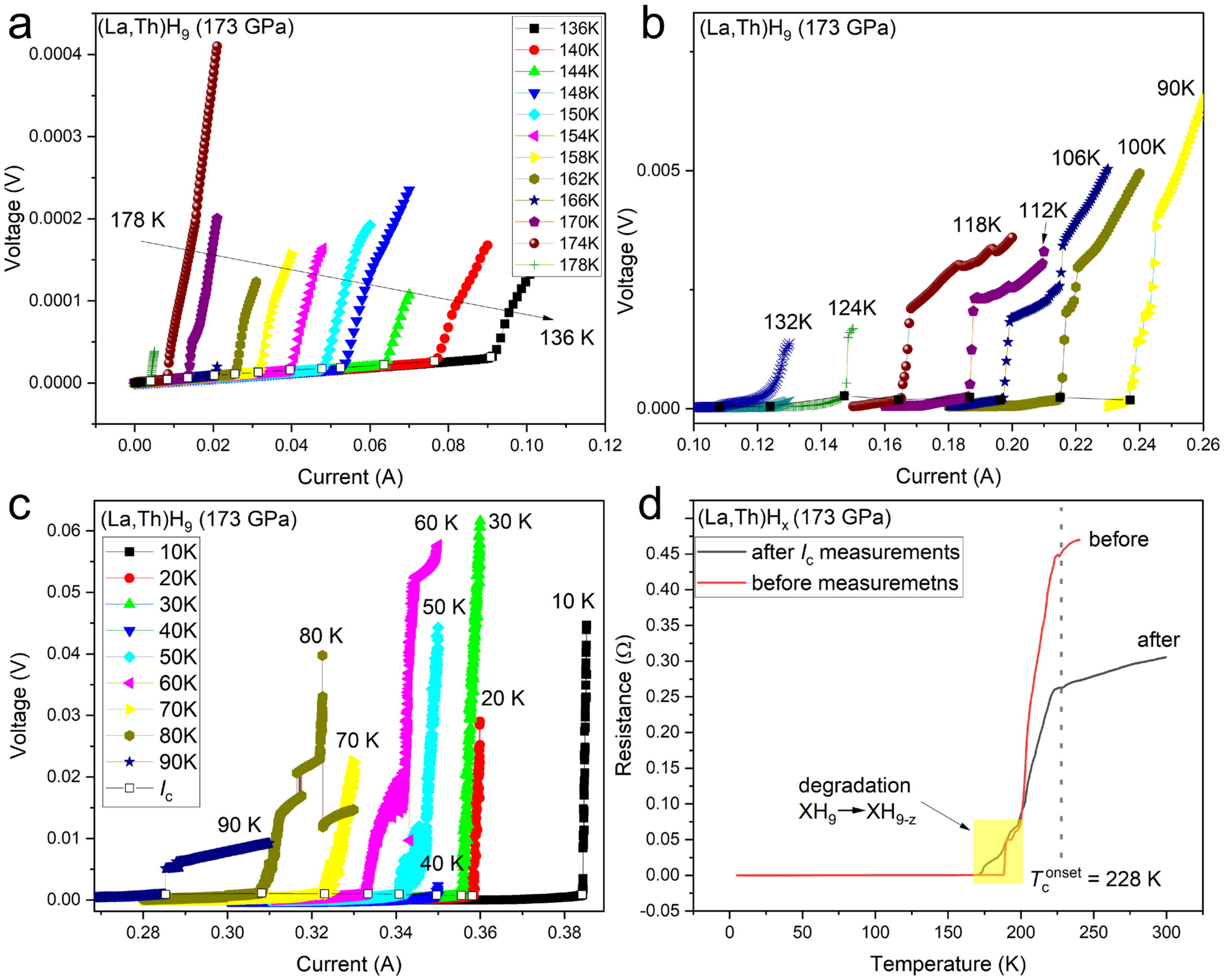

**Figure S24.** Current-voltage characteristics of the (La,Th)$H_9$ bridge in the vicinity of $T_c$(offset) at 173 GPa, measured using different pulsed current sources. (a) Measurements up to 0.1 A using a Keithley 6221a in the temperature range from 178 K to 136 K. (b) Current-voltage characteristic measured in the temperature range from 132 K to 90 K using a custom-made pulsed current source. (c) The same for the range from 90 K to 10 K. (d) Comparison of temperature dependences of electrical resistance before and after critical current measurements, showing the presence of some degradation of the sample (i.e., (La,Th)$H_9$ bridges) due to heating and diffusion of hydrogen atoms and ions under the action of the electric field of the pulsed current.

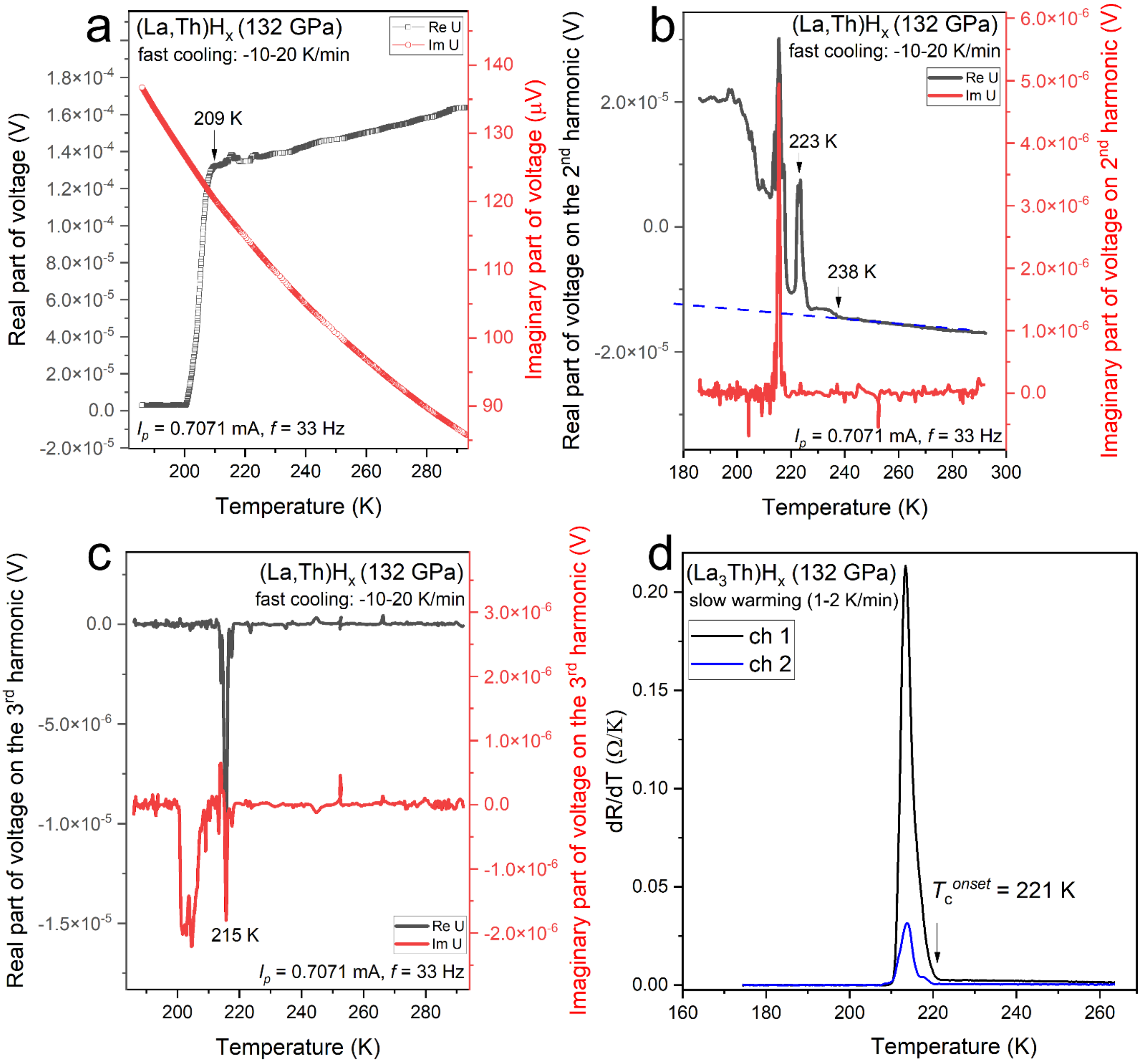

**Figure S25.** Temperature dependence of linear and nonlinear electrical response in (La,Th)$H_{9\text{-}10}$ at 132 GPa in pulsed field DAC loaded with $La_3Th$ alloy. (a) Real (Re U, black) and imaginary (Im U, red) parts of the AC voltage measured under fast cooling conditions (-10 to -20 K/min) at a peak current $I_p$ = 0.7071 mA and frequency $f$ = 33 Hz, showing a distinct transition feature around 209 K. In this situation thermometer shows lower temperature than the real sample's temperature. (b) Real (Re U, black) and imaginary (Im U, red) components of the 2nd harmonic voltage response, highlighting features at 223 K and 238 K. (c) Real (Re U, black) and imaginary (Im U, red) components of the 3rd harmonic voltage response, displaying a prominent signal near 215 K and below. (d) Temperature derivative of resistance (*dR/dT*) measured across two channels (ch 1 and ch 2) during slow warming (1 to 2 K/min).

## 6. Radio-frequency transmission measurements

In the radio-frequency transmission method [22], we detect anomalies of the transmitted signal in the vicinity of the superconducting transition temperature. These anomalies may have different shapes: peaks, steps, or kinks. It is important to note that the finiteness of the superconducting transition width detected in the resistive measurements naturally corresponds to the spread of the temperatures of transmission anomalies in the RF method, detected at different frequencies. Superconducting transitions in individual micro grains of the sample occur at different temperatures, and only combining data for a wide frequency band gives a complete picture of the broadened SC transition, similar to that observed in resistivity measurements.

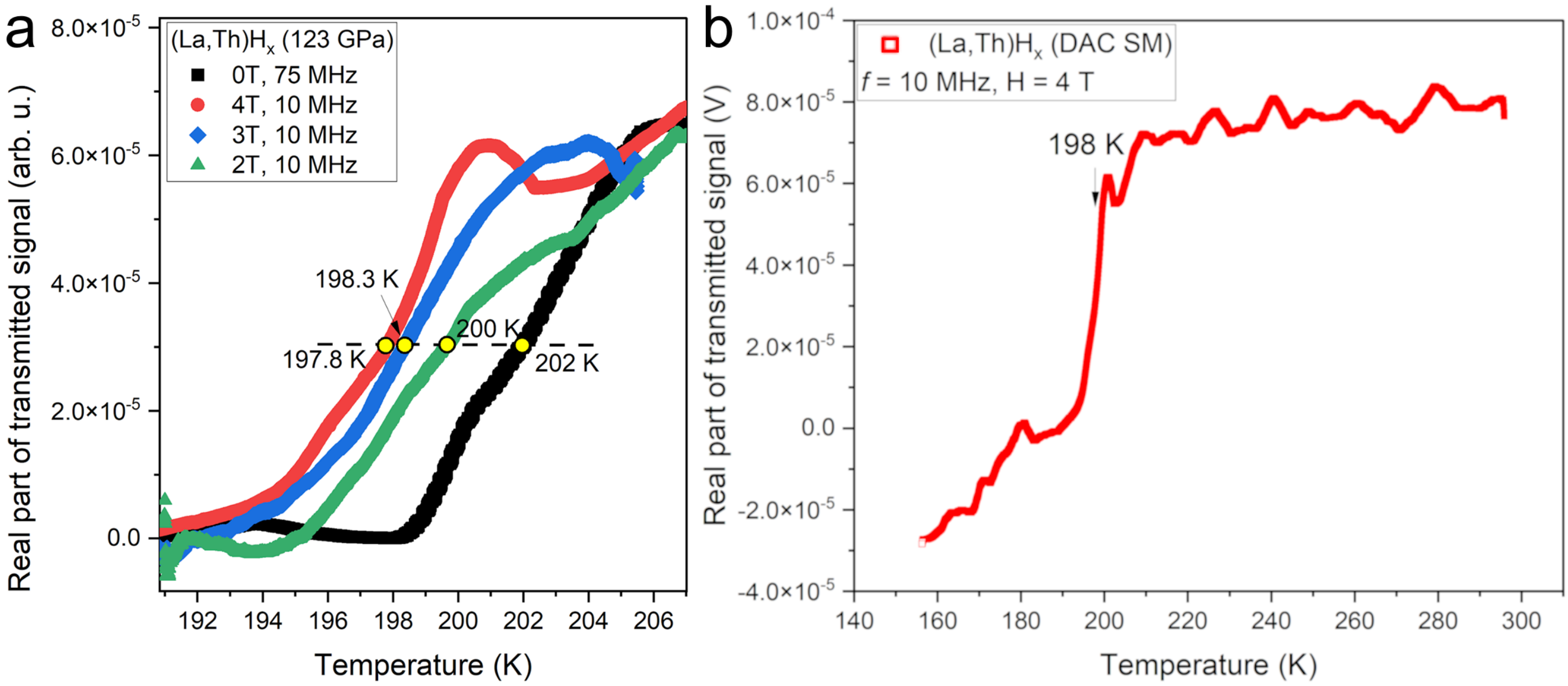

**Figure S26.** Radio-frequency transmittance measurements in DAC SM at 123 GPa in a magnetic field of 0-4 T and an example of the RF signal in the full temperature range. (a) Shift of the superconducting transition signature in the RF channel when an external magnetic field from 0 to 4 T is applied, cooling cycles. $T_c$ was determined by the 50% change in the RF signal amplitude between the bottom and the top of the signal step. (b) Plot of transmitted RF signal (DAC SM) in the full temperature range with a pronounced superconducting transition detected at the carrier frequency $f = 10$ MHz in the magnetic field $H = 4$ T.

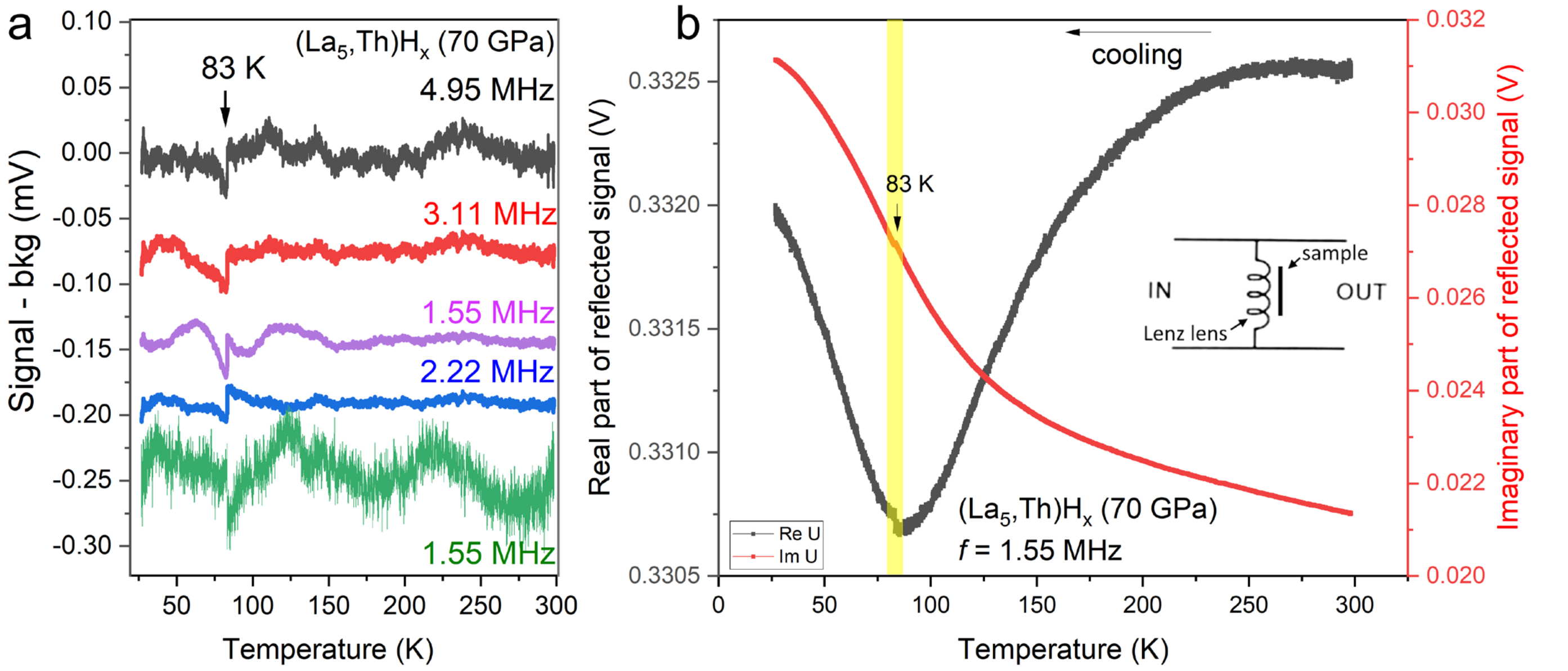

**Figure S27.** Radio-frequency study of DAC S5 with a sample ($La_5$,Th)Hx at 81 GPa in a reflection mode. (a) Dependence of the reflected signal (in mV for black curve, and in arb. units for the rest) on temperature after subtracting the polynomial background at different frequencies. There is a pronounced anomaly at 83 K. (b) Signal type before subtracting the background. The only pronounced anomaly is noticeable at 83 K. Inset: experimental scheme and signal propagation from the output of the generator to the input to the MFLI lock-in amplifier.

Since the choice of frequencies for RF measurements is an ambiguous procedure and requires multiple temperature scans at different frequencies (MFLI frequency range is limited by 5 MHz and 4 generators), a more promising approach seems to be to detect the amplitude of the passing white noise containing all frequencies simultaneously (up to 160 MHz, Hantek HFG2102B generator) as shown in Figure S28a-c. An example of such an integral process is shown in Figure S28d, where the amplitude of the incoming noise was chosen to be 1V, the sample is DAC SM (matrix). At the output, the noise signal was integrated and its power was recorded on a logarithmic scale as a function of temperature. Such a noise detection scheme makes it possible to determine the presence of superconducting phases (La,Th)$H_{9-10}$ in the DAC SM sample with a critical superconducting junction temperature of 193-213 K.

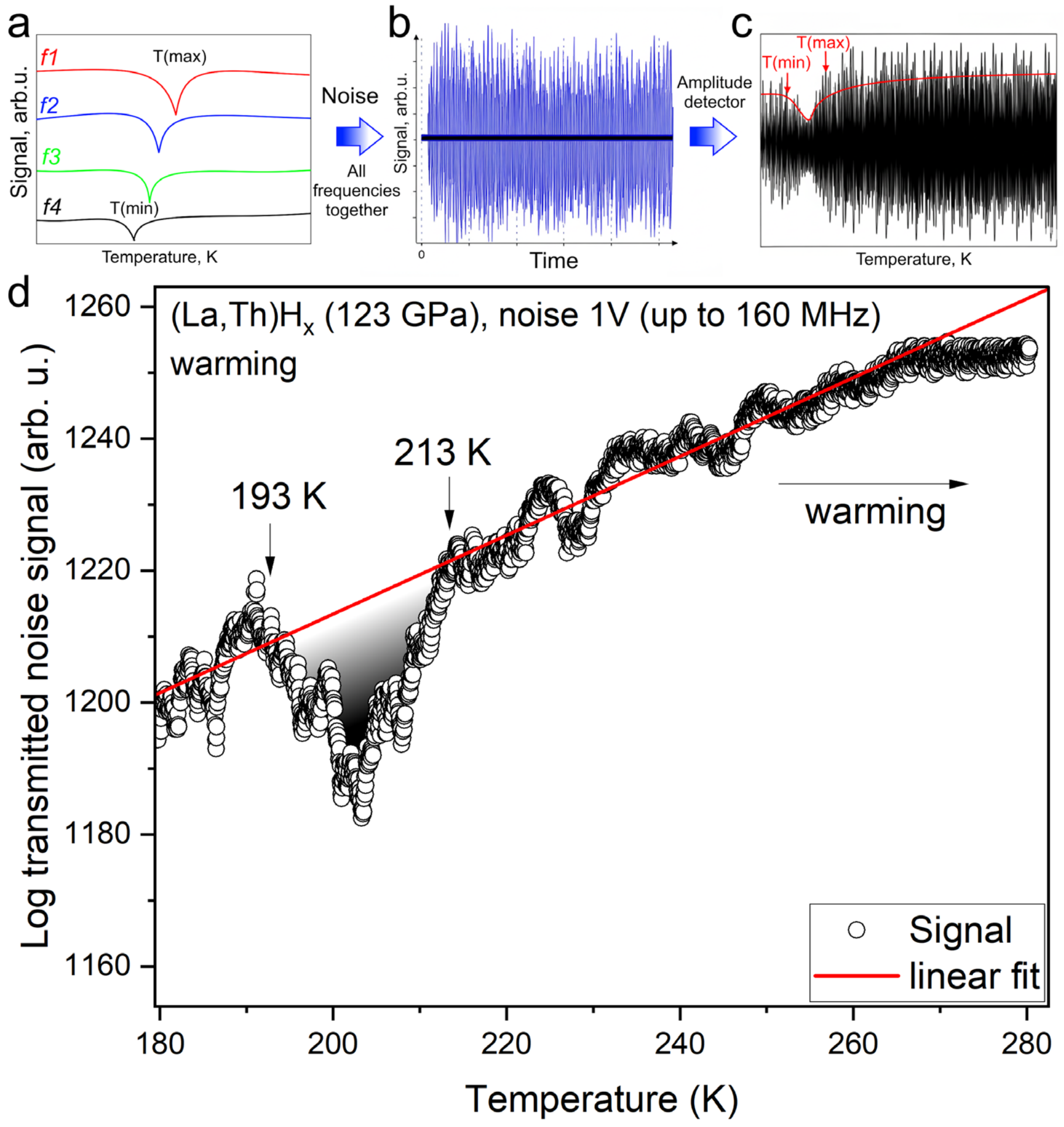

**Figure S28.** Example of measurements with a noise generator. (d) Logarithm of the transmitted noise signal amplitude as a function of temperature measured in the warming cycle for DAC SM with (La,Th)$H_{10}$ at 123 GPa. Noise amplitude was 1V, and the frequency range was up to 160 MHz. The black circles represent the experimental data, and the red line is a linear fit to the normal state region. A distinct feature (drop of the noise level) was observed between 193 K and 213 K, indicating changes in the sample's RF response in a broad frequency range. The top panels (b-c) illustrate the measurement principle: a broadband noise signal is transmitted through the sample, and the amplitude of the transmitted signal is analyzed as a function of temperature. The top-left panel (a) shows representative frequency-dependent transmission RF signals which can be detected at different temperatures at different frequencies.

## 7. Theoretical calculations

Simple estimation calculations of electron-phonon interactions in La-Th-H and Y-Th-H systems using the tetrahedral method, coarse 16×16×12 k-grid and 4×4×2 q-grid indicate a high potential critical temperature of superconductivity, about 185 K at 180 GPa, in the substituted phase $LaThH_{20}$ made from *fcc*-$LaH_{10}$. In addition, theoretical calculations (Figure S29) confirm the following inequalities

$$\mathrm{T_c}(LaThH_{20}) > \mathrm{T_c}(LaThH_{18}) > \mathrm{T_c}(YThH_{18})$$

which are also observed in the experiment [23]. Slightly better grid densities (18×18×18 k-grid and 3×3×3 q-grid) were tested on *hcp*-$ThH_9$ and resulted in a substantial increase in the calculated $T_c$ and improved agreement with the experiment (Figure 29d).

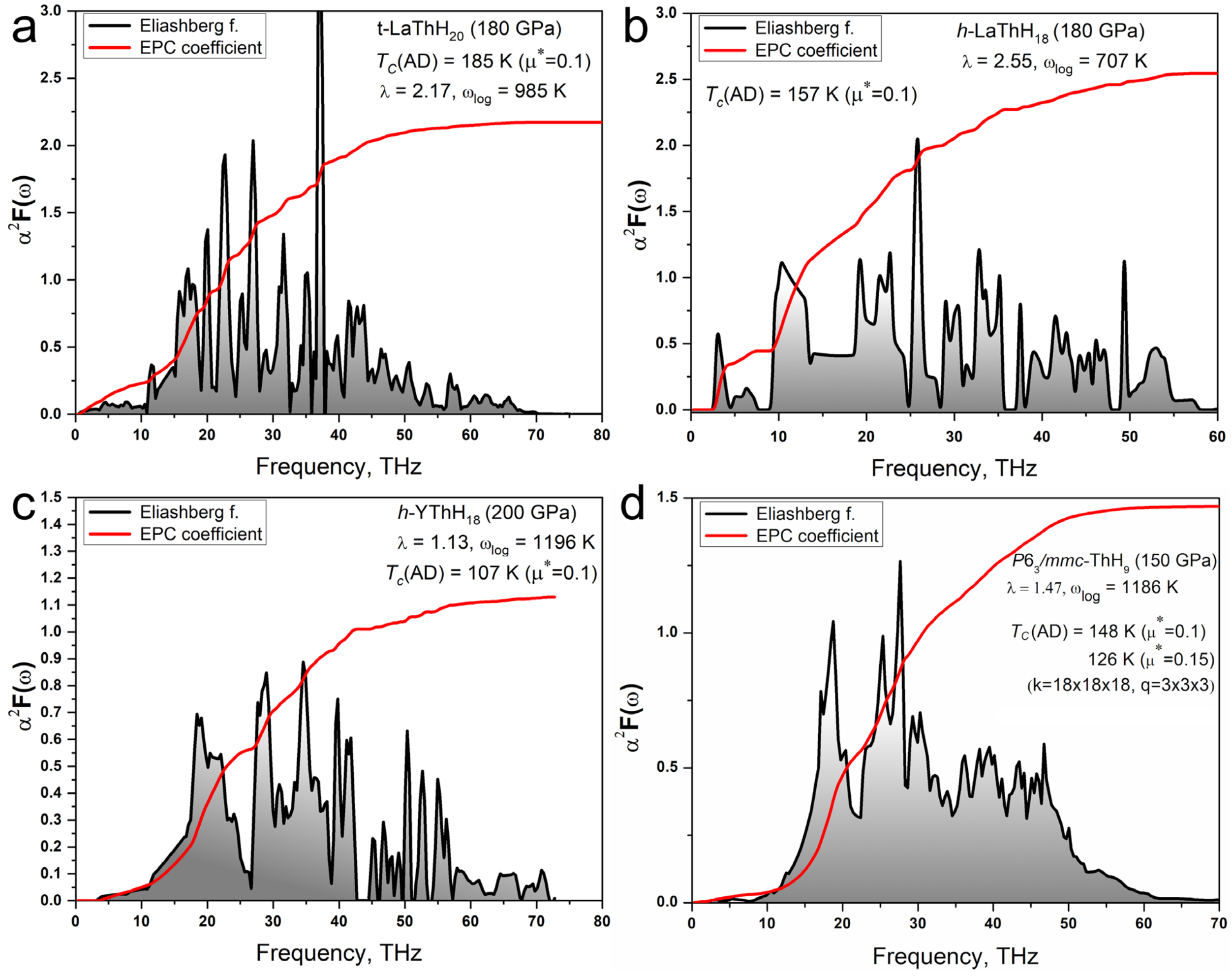

**Figure S29.** Examples of the electron-phonon interaction calculations in La-Th-H and Y-Th-H systems at 180-200 GPa. (a) Eliashberg function and electron-phonon interaction parameters for the product of substitution of half of the La atoms by Th in cubic $Fm\bar{3}m$-$La_4H_{40}$ at 180 GPa. Relaxation of the $La_2Th_2H_{40}$ (= $LaThH_{20}$) compound leads to a change in the lattice symmetry to tetragonal (t). (b) Eliashberg function and electron-phonon interaction parameters for the product of substitution of half of the La atoms by Th in hexagonal $P6_3/mmc$-$La_2H_{18}$ at 180 GPa. (c) Eliashberg function and electron-phonon interaction parameters for the product of substitution of half of the La atoms by Th in hexagonal $P6_3/mmc$-$Y_2H_{18}$ at 200 GPa. (d) Eliashberg function and electron-phonon interaction parameters for the $P6_3/mmc$-$Th_2H_{18}$ at 150 GPa.